\documentclass[12pt]{article}
\usepackage{a4}
\usepackage{amsfonts}
\usepackage{amssymb}
\usepackage{cite}
\usepackage{epsfig}
\usepackage{amsmath}
\usepackage{cite}
\author{}

\newcommand{\drawsquare}[2]{\hbox{%
\rule{#2pt}{#1pt}\hskip-#2pt%  left vertical
\rule{#1pt}{#2pt}\hskip-#1pt%  lower horizontal
\rule[#1pt]{#1pt}{#2pt}}\rule[#1pt]{#2pt}{#2pt}\hskip-#2pt%  upper horizontal
\rule{#2pt}{#1pt}}% right vertical
\newcommand{\fund}{\raisebox{-.5pt}{\drawsquare{6.5}{0.4}}}%  fund
\newcommand{\antifund}{\overline{\fund}}

\newcommand{\be}{\begin{equation}}
\newcommand{\ee}{\end{equation}}
\newcommand{\ba}{\begin{array}}
\newcommand{\ea}{\end{array}}
\newcommand{\bea}{\begin{eqnarray}}
\newcommand{\eea}{\end{eqnarray}}

\newcommand{\ov}{\overline}
\def\IR{\relax{\rm I\kern-.18em R}}

\def\IP{\relax{\rm I\kern-.18em P}}

\def\inbar{\vrule height1.5ex width.4pt depth0pt}
\def\IC{\relax\,\hbox{$\inbar\kern-.3em{\rm C}$}}
\def\K3{{\bf K3}}

\def\ov{\overline}

\def\n2d{\cN_{V^*}^{\otimes 2}}

\def\IC{\mathbb{C}}

\def\IR{\mathbb{R}}

\def\IP{\mathbb{P}}

\def\cN{{\mathcal N}}

\def\to{\rightarrow}

\begin{document}

\title{
\begin{flushright} \vspace{-2cm}
{\small MPP-2007-100\\
 \small UPR-1180-T\\
}
\end{flushright}
\vspace{1.8cm}
 Lifting D-Instanton Zero Modes by  Recombination and Background Fluxes}
\vspace{2.5cm}
\author{\small Ralph~Blumenhagen$^{1}$, Mirjam Cveti{\v c}$^{2}$, Robert Richter$^{2}$   and  Timo Weigand$^{2}$}

\date{}

\maketitle

\begin{center}
\emph{$^{1         }$ Max-Planck-Institut f\"ur Physik, F\"ohringer Ring 6, \\
  80805 M\"unchen, Germany } \\
\vspace{0.1cm}
\emph{$^{2        }$ Department of Physics and Astronomy, University of Pennsylvania, \\
     Philadelphia, PA 19104-6396, USA } \\

\vspace{0.2cm}

\tt{blumenha@mppmu.mpg.de, cvetic@cvetic.hep.upenn.edu, rrichter@sas.upenn.edu, timo@sas.upenn.edu}
\vspace{0.1cm}
\end{center}
\vspace{1.0cm}

\begin{abstract}
\noindent
We study the conditions under which D-brane instantons in Type II orientifold compactifications
 generate a non-perturbative superpotential.
If the instanton is non-invariant under the orientifold action, it
carries four instead of the two Goldstone fermions required for
superpotential contributions. Unless these are lifted, the instanton
can at best generate higher fermionic F-terms of Beasley-Witten
type. We analyse two strategies to lift the additional zero
modes. First we discuss the process of instantonic brane
recombination in Type IIA orientifolds. We show that in some cases
charge invariance of the measure enforces the presence of further
zero modes which, unlike the Goldstinos, cannot be absorbed. In
other cases, the instanton exhibits reparameterisation zero modes
after recombination and a superpotential is generated if these are
lifted by suitable closed or open string couplings. In the second
part of the paper we address  lifting the extra
Goldstinos of $D3$-brane instantons  by supersymmetric three-form
background fluxes in Type IIB orientifolds. This requires
non-trivial gauge flux on the instanton. Only if
the part of the fermionic action linear in the gauge flux survives
the orientifold projection can the extra Goldstinos be lifted.

\end{abstract}

\thispagestyle{empty}
\clearpage

\tableofcontents

\section{Introduction}

Since the recent observation that D-brane instantons in Type II
orientifolds can induce an important new class of effective
couplings
\cite{Blumenhagen:2006xt,Haack:2006cy,Ibanez:2006da,Florea:2006si},
a lot of effort has gone into further exploring these
and other interesting non-perturbative effects
\cite{Abel:2006yk,Akerblom:2006hx,Bianchi:2007fx,Cvetic:2007ku,Argurio:2007qk,Argurio:2007vq,Bianchi:2007wy,Ibanez:2007rs,Akerblom:2007uc,Antusch:2007jd,Yukawas,Aharony:2007pr},
with directly related earlier work including
\cite{Witten:1996bn,Ganor:1996pe}.

In the case of Type IIA orientifolds with intersecting
$D6$-branes, the relevant non-perturbative objects are Euclidean $D2$-brane instantons,
short $E2$-instantons, wrapping
 special Lagrangian three-cycles of the internal Calabi-Yau space \cite{Blumenhagen:2006xt,Ibanez:2006da}.
An analysis of  the
zero mode structure of such instantons can be performed with the help of boundary CFT methods as originally applied to the $D3-D(-1)$ system in \cite{Green:2000ke,Billo:2002hm}. This has shown
that under suitable circumstances the  $E2$-instanton
can generate
couplings in the effective four-dimensional superpotential which are
forbidden perturbatively as a consequence of  global $U(1)$ selection
rules. The relevant instanton effect is genuinely stringy in that it cannot be understood in terms of four-dimensional gauge instantons.

The imprints of this phenomenon in various corners of the string landscape are manifold.
Of particular phenomenological interest has been the generation of
Majorana mass terms for right-handed neutrinos
\cite{Blumenhagen:2006xt,Ibanez:2006da,Cvetic:2007ku,Ibanez:2007rs,Antusch:2007jd}.
Besides allowing for such terms in the first place, instanton effects
admit a natural engineering of the intermediate mass scale
required for these Majorana terms in the context of the see-saw
mechanism. Other applications include the generation of
hierarchically small $\mu$-terms
\cite{Blumenhagen:2006xt,Ibanez:2006da} or a modification of the
family structure of Yukawa couplings \cite{Abel:2006yk}. In
\cite{Yukawas}, the generation of perturbatively forbidden ${\bf
10}\,\,{\bf 10}\,\,{\bf 5_H}$ couplings in SU(5) GUT models based on
intersecting branes is discussed. Globally defined examples of an instanton induced
lifting of unwanted chiral exotics are presented in  \cite{Bianchi:2007fx, Yukawas}. The benefits of instanton effects for realising metastability and supersymmetry breaking in explicit setups are explored in \cite{Florea:2006si,Argurio:2007qk,Aharony:2007pr}.

Using the CFT description for the computation of $E2$-instanton
generated superpotential couplings proposed in
\cite{Blumenhagen:2006xt}, the non-perturbative Majorana mass matrix
for right-handed neutrinos was determined in detail for a local
GUT-like toroidal brane setup in \cite{Cvetic:2007ku}. An extensive
search for realisations of this effect
 within the class
\cite{Anastasopoulos:2006da} of global semi-realistic Gepner model
orientifolds has been performed  in \cite{Ibanez:2007rs}, followed
by further phenomenological studies in \cite{Antusch:2007jd}.

The main obstacle for finding appealing global string vacua exhibiting a non-perturbative superpotential of the described type are the severe restrictions on the zero mode structure of the instanton, which will be reviewed in detail in section \ref{sec_supo} of this article.
At least in the absence of other mechanisms to lift the fermionic zero modes associated with deformations of the cycle, the instanton has to be rigid. Unfortunately, for toroidal orbifolds, a popular playground for Type IIA model building, the only known examples of such cycles are the ones on the ${\mathbb Z}_2 \times {\mathbb Z}_2'$ orbifold analysed in \cite{Dudas:2005jx,Blumenhagen:2005tn,Blumenhagen:2006ab} and used in the local setup of \cite{Cvetic:2007ku}.

A second complication, which is the central topic of this paper,
occurs for $E2$-instantons on non-invariant cycles, called $U(1)$
instantons in the following. It is given by the appearance of four
Goldstino modes $\theta^{\alpha}, \ov \tau^{\dot \alpha}$, $\alpha,
\dot \alpha=1,2$ instead of the two Goldstinos  $\theta^{\alpha}$
required for the generation of a superpotential
\cite{Argurio:2007vq,Bianchi:2007wy,Ibanez:2007rs}. If the instanton
lies on top of an appropriate orientifold plane, the two extra modes
$\ov \tau^{\dot \alpha}$ are projected out and the instanton can
induce a superpotential term. Given its significance  for the
topography of the landscape of string vacua, it is obviously quite
important to investigate if this is actually the only configuration
of D-brane instantons which induces quantum corrections of the
superpotential.

The key point is to decide if there exists a way to lift the two extra Goldstinos $\ov \tau^{\dot\alpha}$ other than by projecting them out.
%In so doing we follow two different strategies.
%We begin with a review of the technical details of instanton calculus.
Generally speaking, this requires contact terms in the instanton moduli action involving the  modes  $\ov \tau^{\dot\alpha}$ such that they can be soaked up in the path integral without giving rise to higher derivative or higher fermionic  terms in the non-perturbative couplings.

We investigate two different strategies to achieve this.
In section \ref{sec_instrec}, we
analyse couplings of the $\ov \tau^{\dot\alpha}$ modes to massless states in the $E2-E2'$ sector, which likewise have to be absorbed. As a consequence of the D-term constraints for the bosonic zero modes the lifting of these modes requires the presence of a non-vanishing Fayet-Iliopoulos term. The latter arises after slightly deforming the background such that the
$\Xi-\Xi'$ pair of instantonic branes recombines.

We describe in detail the zero mode structure of the $U(1)$
instantons and how it  changes
by the process of condensation of the bosonic modes.
We find that for chiral $\Xi-\Xi'$ recombination,
due to charge conservation the recombined object
always contains extra fermionic zero modes which cannot be absorbed
by pulling down either closed string or matter fields.

However, for non-chiral  $\Xi-\Xi'$ recombination one obtains
an $O(1)$ instanton with deformations. In the Type I dual model it corresponds
to an $E1$-instanton which wraps a holomorphic curve moving
in a family as discussed by Beasley and Witten in \cite{Beasley:2005iu}.
We show that such instantons can generate, in addition to the results of  \cite{Beasley:2005iu},  multi-fermion
couplings also for matter field superpotentials
and under certain circumstances can also
contribute to the superpotential.
Independently of the issue of instanton recombination, in the absence of $E2-E2'$ modes the measure of rigid $U(1)$ instantons is just right to generate possibly open string dependent multi-fermion F-terms which correct the metric on the complex structure moduli space. This is the subject of section 4.

%contributions to the K\"ahler potential for the  complex structure moduli. These effects are inherited from the half-BPS instanton corrections of the hypermultiplet metric in the underlying ${\cal N}=2$ theory before orientifolding \cite{Halmagyi:2007wi}.

An alternative mechanism to eliminate the $\ov \tau^{\dot \alpha}$ modes, speculated upon already in the literature \cite{Argurio:2007vq,Ibanez:2007rs,Aharony:2007pr}, consists in turning on supersymmetric background fluxes. The hope would be that in their presence the instanton does not feel the full ${\cal N}=2$ supersymmetry algebra preserved locally away from the orientifold, but only the ${\cal N}=1$ subalgebra preserved by the fluxes. This should then result in only two as opposed to four Goldstinos.

The lifting of reparametrisation zero modes of $M5$-brane or Type IIB $D3$-brane instantons has been studied in detail \cite{Kallosh:2005yu,Saulina:2005ve,Tripathy:2005hv,Bergshoeff:2005yp, Park:2005hj, Lust:2005cu} (see also \cite{Martucci:2005rb,Gomis:2005wc,Marchesano:2006ns}). The analysis consists in determining the bilinear couplings of the fermionic zero modes to the background fluxes responsible for their lifting.
In section \ref{sec_fluxes} we recall, building upon the expressions for the fermion bilinears derived in \cite{Tripathy:2005hv,Bergshoeff:2005yp}, that in Type IIB orientifolds a lifting of the $\ov \tau^{\dot \alpha} $ of $E3$-instantons is not possible as long as one sticks to supersymmetric three-form flux.
As we then show, this generically  continues to hold even for  $E3$-instantons with gauge flux which are mirror symmetric to Type IIA $U(1)$ instantons at general angles. A possible exception are compactifications with divisors allowing for anti-invariant two-cycles. We illustrate this point in a local example and finally summarize our findings in section 6.

\section{Instanton generated F-terms}
\label{sec_supo}

We are interested in ${\cal N}=1$ supersymmetric
Type II orientifold compactifications to four dimensions.
While what we have to say in the sequel applies, {\emph{mutatis mutandis}}, equally well to Type IIA and Type IIB constructions, we focus here for definiteness on the first case.
We will therefore be working in the context of intersecting $D6$-brane models (see \cite{Uranga:2003pz,Kiritsis:2003mc,Lust:2004ks,Blumenhagen:2005mu,Blumenhagen:2006ci,Marchesano:2007de} for reviews).
The relevant spacetime instantons are given by
$E2$-branes wrapping special Lagrangian three-cycles $\Xi$ in the Calabi--Yau so that they
are point-like in four-dimensional
spacetime. Part of the following two subsections \ref{Sec_E2zero} and \ref{Sec_Gensup} reviews some of the findings of  \cite{Blumenhagen:2006xt,Ibanez:2006da}, while in \ref{sec_BW} we discuss higher fermionic F-terms.

%Such instantons can contribute
%to the holomorphic superpotential only
%if they preserve half of the ${\cal N}=1$ supersymmetry.
%This means that the instanton measure must contain a factor
%$d^4 x\, d^2\theta$.

\subsection{$E2$-instanton zero modes}
\label{Sec_E2zero}

There are two kinds of instanton zero modes according to their charge under the gauge groups on the $D6$-branes.

The uncharged zero modes arise from the $E2$-$E2$ sector.
They always comprise the universal four bosonic Goldstone zero modes $x^{\mu}$ due to the breakdown of four-dimensional Poincar{\'e} invariance. Generically, for instantons away from the orientifold fixed plane, these come with four fermionic zero modes $\theta^{\alpha}$ and $\overline \tau^{\dot \alpha}$ \cite{Argurio:2007vq,Bianchi:2007wy,Ibanez:2007rs}. This reflects the fact that the instanton breaks half of the eight supercharges preserved by the Calabi-Yau manifold away from the orientifold fixed plane. Due to its localisation in the four external dimensions, an instanton breaks one half of the ${\cal N}=1$ supersymmetry preserved by the orientifold and one half of its orthogonal complement inside the ${\cal N}=2$ supersymmetry algebra preserved by the Calabi-Yau. As displayed in table \ref{tab_zero}, the $\theta^{\alpha}$  are the Goldstinos associated with the breakdown of the first ${\cal N}=1$ supersymmetry, while the $\overline \tau^{\dot \alpha}$  are associated with the orthogonal  ${\cal N}=1'$ algebra\footnote{Note that what was called $\ov \theta^{\dot \alpha}$ in \cite{Cvetic:2007ku} is now denoted by $\overline \tau^{\dot \alpha}$ to make its spacetime interpretation clearer.}. The internal part of their vertex operator is essentially given by the spectral flow operator of the worldsheet ${\cal N}=(2,2)$  superconformal theory, see eq. (\ref{theta -1/2}) and (\ref{ovtheta -1/2}) in appendix \ref{closed_int}.

\begin{table}[ht]
\centering
\begin{tabular}{|c|c|}
\hline
${\cal N}=1$  & ${\cal N}=1'$   \\
\hline \hline
$    \theta^{\alpha}  $ & $ \tau^{\alpha}  $   \\ \hline
$ {  \ov \theta^{\dot \alpha} } $ & $ {\ov \tau^{\dot \alpha} } $   \\
\hline
%\label{tab_zero}
\end{tabular}
\caption{Universal fermionic zero modes $\theta^{\alpha}, \ov \tau^{\dot \alpha}$ ($\tau^{\alpha}, \ov \theta^{\dot \alpha})$ of an (anti-)instanton associated with the breaking of the ${\cal N}=1$ SUSY algebra preserved by the orientifold and its orthogonal complement ${\cal N}=1'$.
\label{tab_zero} } % \vspace{3mm}
\end{table}

%The vertex operators for these Goldstones and Goldstinos are given by
%\bea
%V^{-1}_{x^{\mu}}(z) &=& x^{\mu} \psi_{\mu}(z)\,\, e^{-{\varphi(z)}},\,\, \nonumber \\
%V^{-{1\over 2}}_{\theta}(z) &=& \theta_{\alpha} S^{\alpha}(z)\,\,  \Sigma^{E2,E2}_{3/8}(z)\,\, e^{-{\varphi(z)\over 2}},\,\, \nonumber \\
%V^{-{1\over 2}}_{\ov \theta}(z) &=& \ov\theta_{\dot \alpha} S^{\dot \alpha}(z)\,\,  \Sigma^{E2,E2}_{3/8}(z)\,\, e^{-{\varphi(z)\over 2}}.\,\,
%\eea
Besides there are $b_1(\Xi)$ complex bosonic zero modes $c_I, I= 1, \ldots, b_1(\Xi)$, related to the deformations and Wilson lines of the $E2$-instanton. Away from the orientifold plane, each of these is accompanied by one chiral and one anti-chiral Weyl spinor, $\chi^{\alpha}_I$ and $\ov \chi^{\dot \alpha}_I$.
Furthermore there arise zero modes at non-trivial intersections of the instanton $E2$ with its image $E2'$; they will be discussed in detail in section \ref{sec_zero_U1}.

In addition to these uncharged zero modes, there can arise fermionic zero modes from intersections of  the instanton $\Xi$ with $D6$-branes $\Pi_a$.  If the instanton is parallel to  $\Pi_a$, there are also massless bosonic modes in this sector.
The detailed quantisation of these charged zero modes, both for chiral and non-chiral intersections, is described in \cite{Cvetic:2007ku}. Let us focus for brevity on chiral intersections.
An important point made in \cite{Cvetic:2007ku} is that states in the $E2-D6$ sector are odd under the GSO projection contrary  to the GSO-even states in the $D6-D6$ brane sector. In particular, a positive intersection $I_{\Xi a} > 0$ of the instanton and a $D6$-brane wrapping the respective
cycles $\Xi$ and $\Pi_a$ hosts a  single
\emph{chiral} fermion (i.e. with world-sheet charge
$Q_{\rm ws}=-\frac{1}{2}$) in the bifundamental representation $(-1_E,\fund_a)$. The strict chirality of the charged fermions is essential for the existence of holomorphic couplings between these modes and open string states in the moduli action and will also play a key role in the present analysis.
For a generic instanton cycle $\Xi$ away from the orientifold, this gives rise to the charged zero mode spectrum summarised in table \ref{tablezero}.
 \begin{table}[ht]
\centering
\begin{tabular}{|c|c|c|}
\hline
zero modes&  Reps$_{Q_{ws}}$ & number   \\
\hline \hline
$\lambda_{a,I}$ &  $(-1_E,\fund_a)_{-1/2}$   & $I=1,\dots, [\Xi\cap \Pi_a]^+$    \\
$\overline{\lambda}_{a,I}$ &  $(1_E,\antifund_a)_{-1/2}$  & $I=1,\dots, [\Xi\cap \Pi_a]^-$    \\
\hline
$\lambda_{a',I}$ &  $(-1_E,\antifund_a)_{-1/2}$ & $I=1,\dots, [\Xi\cap \Pi'_a]^+$    \\
$\overline{\lambda}_{a',I}$    &  $(1_E,\fund_a)_{-1/2}$   & $I=1,\dots,[\Xi\cap \Pi'_a]^-$    \\
\hline
\end{tabular}
\caption{Zero modes at chiral $E2$ -$D6$ intersections.
\label{tablezero} } % \vspace{3mm}
\end{table}
As a result, the instanton carries the charge \cite{Blumenhagen:2006xt,Ibanez:2006da}.
\bea
\label{chargee}
          Q_a(E2)={\cal N}_a\,\, \Xi\circ (\Pi_a - \Pi'_a)
\eea
under the gauge group $U(1)_a$.

\subsection{Generation of superpotentials}
\label{Sec_Gensup}
The instanton measure contains all these zero modes. Thus in order to contribute to the holomorphic superpotential, whose measure is $\int d^4 x\, d^2\theta$, the instanton has to meet several constraints.

Most importantly, the presence of the anti-chiral Goldstinos $\ov
\tau^{\dot \alpha}$ for generic instantons not invariant under the orientifold
projection\footnote{These will be referred to as $U(1)$ instantons
in the sequel.} prevents the generation of superpotential terms
other than those corresponding to gauge instantons
\cite{Argurio:2007vq,Bianchi:2007wy,Ibanez:2007rs}. The latter case
is special in that the instanton wraps the same three-cycle as one
of the $D6$-branes \cite{Akerblom:2006hx}. In this situation, the
$\ov \tau^{\dot \alpha}$ play the role of Lagrange multipliers for the bosonic
ADHM constraints and can consistently be integrated out
\cite{Billo:2002hm}. For instantons not parallel to any of the
$D6$-branes, these couplings in the moduli action do not exist since
there are no massless bosons in the $E2-D6$ sector. The most
straightforward way to eliminate the $\ov \tau^{\dot \alpha}$ is to project them
out under the orientifold action
\cite{Argurio:2007vq,Bianchi:2007wy,Ibanez:2007rs}. Concretely, if
one chooses $\Xi=\Xi'$  the universal zero modes $x^{\mu}, \theta^{\alpha},
\ov \tau^{\dot \alpha}$ are subject to the orientifold action $\Omega\ov\sigma$ in
the way detailed in appendix \ref{orientifold}. Depending on the
orientifold action one obtains an $SO(N)$ or $USp(N)$ gauge group.
For the latter case the zero modes $x^\mu,\theta^{\alpha}$ are anti-symmetrised
and the modes $\ov\tau^{\dot \alpha}$ gets symmetrised, while for the $SO(N)$
instanton $x^\mu,\theta^{\alpha}$ are symmetrised and $\ov\tau^{\dot \alpha}$ get
anti-symmetrised.

It follows that single $E2$-instantons with orthogonal gauge group (called $O(1)$ instantons in the sequel) can give  rise to F-terms in the effective action since the universal part of their zero mode measure is of the form $\int d^4 x \, d^2 \theta$.
%Note that for invariant instantons there are no zero modes arising from the $E2$-$E2'$ sector.

In order for this F-term to be of the usual superpotential form, there may be no further
uncharged fermionic zero modes present. This situation corresponds to an instanton wrapping a rigid cycle $\Xi$ with  $b_1(\Xi)=0$.
%The easiest reason for the effective absence of the reparametrisation moduli is of course rigidity of the instanton ($b_1(\Xi)=0$), which avoids the appearance of additional zero modes in the $E2$-$E2$ sector apart from the universal ones.
Alternatively, the additional fermionic modes have to be absorbed by some interaction in the instanton moduli action such that they can be integrated out without generating higher derivative terms.
%, possibly leading to a non-trivial integral over the bosonic moduli space.
Known examples of such interactions involving the closed string
sector are the quartic coupling to the curvature on the instanton
moduli space \cite{Dine:1987bq,Dorey:2002ik}, provided the latter is
non-trivial, or the coupling to suitable background fluxes (see
section \ref{sec_fluxes}). In section \ref{sec_super?} we will
describe another way to lift a pair of reparametrisation modes
through couplings to the open string sector.

Finally, also the charged zero modes appear in the measure and have to be soaked up.
For an $\Omega\ov\sigma$ invariant instanton, i.e. $\Xi'=\Xi$,
the charged zero modes and their representations  are displayed in Table \ref{tablezeroinv}
\begin{table}[ht]
\centering
\begin{tabular}{|c|c|c|}
\hline
zero modes&  Reps. & number   \\
\hline \hline
$\lambda_{a,I}$ &  $\fund_a$   & $I=1,\dots, [\Xi\cap \Pi_a]^+$    \\
$\overline{\lambda}_{a,I}$ &  $\antifund_a$  & $I=1,\dots, [\Xi\cap \Pi_a]^-$\\
\hline
\end{tabular}
\caption{Zero modes at $E2-D6$ intersections.
\label{tablezeroinv} } % \vspace{3mm}
\end{table}
and lead to an instanton $U(1)_a$ charge
\bea
\label{charge_inv}
          Q_a(E2)={\cal N}_a\,\, \Xi\circ \Pi_a.
\eea
A careful analysis of their $g_s$ scaling in \cite{Blumenhagen:2006xt, Cvetic:2007ku} revealed that for superpotential couplings this has to happen via suitable disk (as opposed to higher genus) amplitudes involving precisely two $\lambda$ modes and in addition suitable matter fields  - provided these amplitudes induce a Yukawa-type contact term in the instanton moduli action.
As a result, $E2$-instantons induce  superpotential terms of the form \cite{Blumenhagen:2006xt,Ibanez:2006da}.
\bea
\label{chirsuper}
            W\simeq  \prod_{i=1}^M  \Phi_{a_i,b_i}\, e^{-S_{E2}},
\eea
involving suitable products of open string fields $\Phi_{a_i,b_i}$. For details of the rules of their computation see \cite{Blumenhagen:2006xt}.

\subsection{Generation of higher fermionic F-terms}
\label{sec_BW}

Our discussion has hitherto focussed on $O(1)$ instantons which are either rigid or whose fermionic  reparametrisation modes have paired up appropriately such that they give rise to genuine superpotential terms.

Alternatively, there are situations where these additional zero modes induce so-called higher fermionic F-term couplings in the effective action.
In the dual Type I/heterotic model this effect was first described in \cite{Beasley:2005iu}\footnote{For another example in the context of heterotic M-theory see \cite{Buchbinder:2006xh}.}. There it arises for
$E1$/worldsheet instantons moving in a family. On the type IIA side, this corresponds to non-rigid $O(1)$ instantons such that the chiral reparametrisation modulini $\chi_I^{\alpha}, I=1, \ldots, b_1(\Xi)$ are anti-symmetrised and therefore projected out under the orientifold action. We will sometimes refer to them as instantons with deformations of the first kind\footnote{This is to be contrasted with the case that the chiral deformation fermions survive the projection. As descrbed in \cite{Akerblom:2007uc} such a situation can generate corrections to the gauge kinetic function.}.
The resulting uncharged part of the measure takes the form
\bea
\label{measure_family}
\int d^4 x \, d^2 \theta \, \prod_I c_I \, \ov c_I \, \ov \chi_I^{\dot\alpha}.
\eea

Beasley and Witten found that such instantons can generate
higher fermionic couplings for the closed string moduli
fields \cite{Beasley:2005iu}.
In superspace notation, these are encapsulated in interactions of the form
\bea
S = \int d^4 x \, d^2 \theta \, \, w_{\ov i \ov j}\, (\Phi) {\ov{\cal D}}^{\dot \alpha} {\ov \Phi}^{\ov i}  {\ov{\cal D}}_{\dot \alpha} {\ov \Phi}^{\ov j}
\eea
for the simplest case that the instanton moves in a one-dimensional moduli space.
Note that supersymmetry requires a holomorphic dependence of $w_{\ov i \ov j}\, (\Phi)$ on the superfields $\Phi$.

Consider first the case of an $E2$-instanton with $b_1(\Xi) =1$ and no further charged zero modes in the $E2-D6$ sector. Denoting by ${\cal T} = T + \theta^{\alpha} t_{\alpha} $ the ${\cal N}=1$ chiral superfield associated with the K\"ahler moduli, we can absorb the instanton modulini by pulling down from the moduli action two copies of the schematic anti-holomorphic coupling $\ov \chi^{\dot \alpha} \ov t_{\dot \alpha}$.
In general the open-closed amplitude $\langle \ov \chi^{\dot \alpha} \ov t_{\dot \alpha} \rangle$ does not violate any obvious selection rule of the ${\cal N}=(2,2)$ worldsheet theory and is therefore expected to induce the above coupling\footnote{In particular, the {\it total} U(1) worldsheet charge is conserved. Still there might be situations, such as factorizable 3-cycles on $(T^2)^3$, where some of the {\it individual} U(1) charges are violated by this coupling. For a generic background, though, the couplings need not be vanishing, as we demonstrate for the example of a non-factorizable $T^6$ in  appendix \ref{closed_int}.}.
Similarly, the two $\theta$-modes can be soaked up by the holomorphic coupling  $\theta^{\alpha} u_{\alpha}$ involving the fermionic partners of the complex structure moduli encoded in the superfield ${\cal U} = U + \theta^{\alpha} u_{\alpha} $. This results in a four-fermion interaction of the schematic form
$e^{-S_{E2}} \, u^{\alpha} u_{\alpha} \, \ov t^{\dot \alpha} \ov t_{\dot \alpha}$. Note that the coupling of the complex and K\"ahler structure modulini only to the universal and reparametrisation zero modes, respectively, is a consequence of  $U(1)$ worldsheet charges of the  associated vertex operators.

The derivative superpartner of the above four-fermi term arises upon integrating out two copies of the term
\bea
\theta \sigma^{\mu} \ov \chi  \, \partial_{\mu} \ov T,
\eea
which follows from evaluating the amplitude $\langle \theta^{\alpha} \,\, \ov {\chi}^{\dot \alpha} \,\, \ov T \rangle$ as demonstrated in appendix \ref{closed_int}.
All this can be summarized in superspace notation by writing
\bea
\label{BW_2}
S = \int d^4 x \, d^2 \theta \, \, e^{-{\cal U}(\Xi)} \,\, f_{\ov i, \ov j}\left(e^{{\cal T}_i}, e^{\Delta_i}\right) {\ov{\cal D}}^{\dot \alpha} {\ov {\cal T}}^{\ov i}  {\ov{\cal D}}_{\dot \alpha} {\ov {\cal T}}^{\ov j},
\eea
where ${\cal U}(\Xi)$ is associated with the specific combination of complex structure moduli appearing in the classial instanton action and the holomorphic function $f_{\ov i, \ov j}$ depends in general on the K\"ahler and open string moduli of the $D6$-branes $\Delta_i$.

In the presence of a suitable number of charged $\lambda$ zero-modes there exist, in addition to these closed string couplings, terms which
generate higher fermi-couplings also for the matter fields.
Consider again for simplicity the case $b_1(\Xi) =1$.
If the Chan-Paton factors and worldsheet selection rules only allow the $\lambda$ modes to couple holomorphically to the chiral open string superfields, as for the generation  of a superpotential,
the instanton induces an interaction as in (\ref{BW_2}), but with $ e^{-{\cal U}(\Xi)}$ replaced by  $ e^{-{\cal U}(\Xi)} \, \prod_{a_i,b_i} \Phi_{a_i,b_i}$.

\begin{figure}[h]
\begin{center}
 \includegraphics[width=0.8\textwidth]{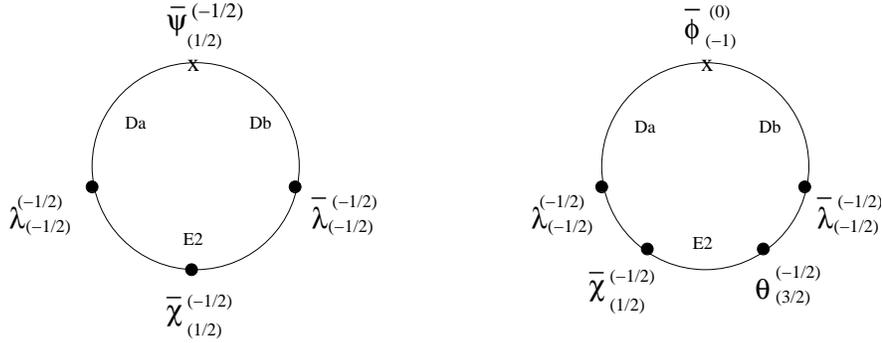}
\end{center}
\caption{\small Absorbtion of $\theta$ and $\bar{\chi}$-modes
leading to F-terms. The superscripts denote the ghost picture. }\label{fermifiga}
\end{figure}

For suitable configurations, the action can also pick up derivative
terms directly involving the open string fields. For this to happen
the instanton moduli action has to contain couplings of the
form\footnote{The subscripts denote the worldsheet $U(1)$-charges,
which are obviously conserved. The actual presence of this contact
term can easily be checked, in the context of toroidal orbifolds, by
a computation analogous to those performed in \cite{Cvetic:2007ku}.}
\bea
 \ov {\chi}^{\dot \alpha}_{1/2}  \lambda^a_{-1/2}\,\,
            (\ov\psi_{1/2})_{\dot\alpha}\,\, \ov \lambda^b_{-1/2},
\eea
where the fermionic matter field $\ov\psi_{1/2}^{\dot\alpha}$
lives at the intersection $D6_a-D6_b$ and lies in the anti-chiral
superfield $\ov\Phi=\ov \phi + \ov\tau \ov\psi$, see figure \ref{fermifiga}.

Integrating out two copies of this interaction term brings down the fermion bilinear $\ov\psi_{1/2}\ov\psi_{1/2} $. %characteristic for the higher fermionic terms described in      \cite{Beasley:2005iu}.
In addition, the two $\theta^{\alpha}$ modes  again pull down a bilinear of {\emph {chiral}} fermions $u^{\alpha}$ or, in the presence of more $\lambda$ modes, $\psi_{ab}^{\alpha}$, as in the case of superpotential contributions. This induces again a four-fermi coupling.
Alternatively, we can absorb one pair of $\theta^{\alpha} \ov\chi^{\dot\alpha}$ in a coupling of the form
\bea
\theta^{\alpha}_{3/2} \,\, \ov {\chi}^{\dot \alpha}_{1/2} \,\,  \lambda^a_{-1/2}\,\, \ov \phi_{-1} \,\, \ov \lambda^b_{-1/2}.
\eea
After bringing the $\ov \phi_{-1}$ into the zero ghost picture this clearly generates a derivative coupling of the form $ \theta \sigma^{\mu} \ov \chi  \,\lambda^a\, \partial_{\mu} \ov \phi \, \ov \lambda^b $. Integrating out two copies of this term yields the derivative superpartner to the above four-fermi term.

\section{Instanton recombination}
\label{sec_instrec}
%In order to completely understand the origin of instanton
%contributions to the superpotential, it is very important
%to know which instantons can contribute.
As just reviewed, for the case of $E2$-instantons in Type IIA orientifolds
we know that single instantons wrapping rigid special Lagrangian
three-cycles invariant under the orientifold
projection and carrying $O(1)$ gauge group
have the right zero mode structure $\int d^4x\, d^2\theta$  to contribute to the superpotential.
Under mirror symmetry to the Type I string these
objects are mapped to $E1$-instantons wrapping
isolated curves on the mirror Calabi-Yau. The contribution
of such objects to the superpotential has been discussed
in a couple of papers \cite{Witten:1999eg,Beasley:2005iu}.

For $D6$-branes it is known that under certain circumstances a pair
of $D6$-$D6'$ branes can recombine \cite{Kachru:1999vj} into a new sLag $D6$-brane  which
obviously wraps an $\Omega\ov\sigma$ invariant three-cycle\footnote{For brane recombination in the context of $D6$-brane model building see e.g. \cite{Cvetic:2001nr,Cremades:2002cs,Blumenhagen:2002gw}.}.  If a
similar story also applies to pairs of $E2$-$E2'$ instantonic
branes, the recombined objects would be candidates for new
$O(1)$-instantons contributing to the superpotential. For example if
one starts with an $E2$-instanton wrapping a factorizable cycle on a
toroidal orbifold, the cycle wrapped by the recombined instanton
would no longer be factorizable; still one could hope to determine
the instanton contribution by appropriate deformation of the
original instanton moduli action. In the mirror dual situation, the
resulting objects are  $E5$-instantons equipped with a vector bundle
$W$ defined via the non-trivial extension \bea \label{extens}
     0\to L \to W\to L^* \to 0
\eea
of the two line bundles  $L$ and $L^*$.

In this section we investigate whether the naive expectation that
such recombined $O(1)$-instantons exist is actually correct.

\subsection{Zero mode structure on $U(1)$ instantons}
\label{sec_zero_U1}

Consider a $U(1)$-instanton wrapping a  general rigid cycle
$\Xi \neq \Xi'$.
From the  $E2-E2$ and $E2'-E2'$ sectors we now have the zero mode measure
\bea
\label{unchargedmeasure}
\int d^4 x\, d^2\theta\, d^2 \ov\tau.
\eea
As described in the previous section, if such an instanton also intersects the $D6$-branes
present in the model, this yields the fermionic zero-modes
 listed in table \ref{tablezero}.
From there, the overall $U(1)_E$ charge of these matter zero modes can be read off,
\bea
\label{chargetot}
\sum_i Q_E(\lambda^i) &=& \sum_a N_a \, \left( -(\Xi \cap \Pi_a)^+ + (\Xi \cap \Pi_a)^-  - (\Xi \cap \Pi_{a'})^+ + (\Xi \cap \Pi_{a'})^- \right)  \nonumber \\
&=& - \sum_a N_a \,  \Xi \circ (\Pi_a + \Pi_{a'}) = - 4 \,\, \Xi
\circ \Pi_{O6} . \eea In the last line we have used the  tadpole
cancellation condition\footnote{Notice that $ \Pi_{O6}$ denotes the
total homological charge of all orientifold fixed planes present in
the background. In what follows we will always refer to the
effective orientifold projection which arises after taking into
account the contribution from all different sectors, which may be of
different types individually. }. This shows that in a globally
consistent model  the total $U(1)_E$ charge of all matter zero modes
is proportional to the chiral intersection between the instanton and
the orientifold plane. For an $\Omega\ov\sigma$ invariant instanton
this last quantity vanishes, whereas for a generic $U(1)$ instanton
it does not.

If $\Xi \circ \Pi_{O6}\ne 0$, there must be additional charged
zero modes in order for the zero mode measure to be $U(1)_E$ invariant.
 Indeed there are also zero modes from the $E2-E2'$ intersection.
This is the open string sector which is invariant under
$\Omega\ov\sigma$ and gets symmetrized or anti-symmetrized (see appendix \ref{orientifold})
.
Taking into account that the sign of the orientifold projection
changes from $Dp$-$Dp$ to $D(p-4)$-$D(p-4)$ sectors,
for a single $U(1)$ instanton we get the zero modes shown in Table \ref{antizero}.

\begin{table}[h]
\centering
\begin{tabular}{|c|c|c|}
\hline
 zero mode & $(Q_E)_{Q_{ws}}$ &   Multiplicity \\
\hline \hline
 $m, \ov m$ & $(2)_1$ ,$(-2)_{-1}$ & ${1\over 2}\left(\Xi'\circ \Xi+\Pi_{{\rm O}6}
\circ \Xi\right)$  \\
 $\ov\mu^{\dot \alpha}$ & $(-2)_{1/2}$ & ${1\over 2}\left(\Xi'\circ \Xi+\Pi_{{\rm O}6}
\circ \Xi\right)$  \\
 $\mu^{\alpha}$ & $(2)_{-1/2}$ & ${1\over 2}\left(\Xi'\circ \Xi-\Pi_{{\rm O}6}
\circ \Xi\right)$  \\
\hline
\end{tabular}
\caption{Charged zero modes at an $E2-E2'$ intersection. % \vspace{3mm}
\label{antizero} }
\end{table}

For concreteness we consider from now on the two simplest
non-trivial cases.

%\vspace{0.2cm}
\noindent
\underbar{Case I}

\noindent
The first case has   intersection numbers
\bea
\label{int_pattern1}
\Xi' \circ \Xi = \Pi_{O6} \circ \Xi=1.
\eea
It corresponds to a projection as would arise e.g. on $T^6 / {\mathbb Z}_2$ in the presence of a single $O^-$-plane.
We get two additional bosonic zero modes
$m$ and $\ov m$ and two additional
fermionic ones $\ov\mu^{\dot\alpha}$.
Comparing with (\ref{chargetot}),  we find that indeed the
total $U(1)_E$ charge of zero modes vanishes. The charge
of the two $\ov\mu^{\dot\alpha}$ zero modes precisely cancels
against the sum over all matter field zero modes.

This analysis reveals that in a globally consistent model it is not
possible to wrap
an $E2$-instanton on a cycle $\Xi \neq \Xi'$ without picking up
additional charged zero modes  $\lambda_i$.
Their $U(1)_E$ charge is guaranteed to cancel the
$U(1)_E$ charge of the $E-E'$ modes such that the resulting zero mode
measure\footnote{Note the inverse scaling behaviour of the Grassmann numbers.},
\bea
\label{measure}
\int d{\cal M}_I = \int d^4 x\,  d^2 \theta d^2\,  \ov \tau \,\,  dm \, \, d \ov m  \,
\, \underbrace{d^2 \overline \mu^{\dot \alpha}}_{Q_E = 4} \, \,
\underbrace{ \prod_b d \ov\lambda_b}_{Q_E = -4}
\eea
 is $U(1)_E$ invariant.

\vspace{0.2cm}
\noindent
\underbar{Case II}

\noindent
The second case has   intersection numbers
\bea
\label{int_pattern2}
\Xi' \circ \Xi = 1, \ \Pi_{O6} \circ \Xi=-1.
\eea
Here  we get no extra bosonic zero modes and only the
two fermionic ones $\mu^{\alpha}$. Unlike the previous case, this is due to a projection as would arise e.g. in the presence of a single $O^+$-plane. In such a situation it is not possible to cancel the tadpoles in a supersymmetric way. Nonetheless, we can perform a similar zero mode analysis.
Again the condition (\ref{chargetot}) tells us that there are
extra fermionic matter zero modes whose $U(1)_E$ charge
is equal to $Q_E=-4$. The resulting zero mode measure reads
\bea
\label{measureb}
\int d{\cal M}_{II} = \int d^4 x\, d^2 \theta \,\, d^2 \ov \tau \,\,
\, \underbrace{d^2 \mu^{\alpha}}_{Q_E = -4} \, \,
\underbrace{ \prod_a d \lambda_a}_{Q_E = 4}.
\eea

\subsection{Recombination of chiral  $E2-E2'$ instantons }
\label{Recom_chiral}

The question we would like to address now is whether one can absorb
the zero modes for the $U(1)$ instantons  in such a way that contributions to the superpotential
$W$ are generated.
The expectation that this might be the case arises
from the analogous situation for intersecting $D6$-branes,
where a slight deformation of the complex structure moduli
induces a non-vanishing Fayet-Iliopolous term on the
$D6$-worldvolume  leading to condensation of the tachyonic
charged matter fields \cite{Kachru:1999vj}. This brane recombination process
preserves the topological charge of the intersecting
$D6-D6'$ branes and therefore yields a supersymmetric brane
wrapping a three-cycle which is invariant under $\Omega\ov\sigma$.

Consider first the {\bf case I} from the last section.
Here we have the bosonic zero modes $m$ and $\ov m$,
which appear in a D-term potential of the form
\bea
\label{rec-action}
S_{E2}= ( 2 m \, \ov m - \xi ) ^2 .
%+ m \, \ov \theta \,\ov \mu
\eea
The complex structure dependent Fayet-Iliopoulos  parameter $\xi$
is proportional to the angle modulo 2 between the cycle $\Xi$ and its
image $\Xi'$ and vanishes for supersymmetric configurations.
Starting from a supersymmetric situation with  $\xi = 0$ one can always
deform the complex structure to obtain $\xi <0 $ or $\xi > 0$
at least for small $\xi$.
Since the geometry of the internal cycle is independent of whether it is wrapped by a $D6$-brane or an $E2$-instanton, we argue that the FI term is forced upon us even in the absence of four-dimensional dynamical gauge fields associated with the abelian gauge group on the instanton.

The D-term constraint resulting from (\ref{rec-action}) is
\bea
m \,\ov m = \frac{1}{2} \, \xi
\eea
and has to be implemented by a delta function in the instanton measure.
It is useful to parametrise the complex boson $m$ as
\bea
m = |m| \, e^{i \alpha}.
\eea
Note that the D-flatness condition as such does not constrain the phase $\alpha$.
The latter can be absorbed by fixing the gauge with respect to the $U(1)_E$ symmetry under which the instanton measure (\ref{measure}) is invariant.

\noindent It follows that the bosonic part of the instanton measure takes the form
\bea
\int d^4 x \, \,  d|m| \, |m|\,  \delta\left(|m|^2 - {\xi}/{2}\right).
\eea
As $\xi$ becomes positive, the bosonic $m$ modes get tachyonic, signalling
an instability towards condensation of the tachyon such that the
D-term constraint is satisfied.
In the \emph{upstairs} geometry
this corresponds to recombination of the cycle $\Xi \cup \Xi'$ (recall that
upstairs $\Xi$ and $\Xi'$ are not identified) to the \emph{unique} sLag $\widetilde \Xi$ with homology class equal to $[\Xi] + [\Xi']$ and $\xi_{new}
= 0$ \cite{Joyce:1999tz}.  Note that $\widetilde \Xi$ is rigid if $\Xi$ (and $\Xi'$)
is rigid \cite{Joyce:1999tz}, i.e. the instanton wrapping it exhibits no uncharged
zero modes apart from the universal ones.

Now we have to determine what happens to the fermionic
zero modes once the bosonic ones condense.
As our analysis of the relevant amplitudes in appendix \ref{closed_int} shows, the instanton moduli action contains the term
\bea
\label{rec-actionb}
S_{E2}=  m \, \ov \tau_{\dot \alpha} \,\ov \mu^{\dot \alpha},
\eea
which means that after $m$ gets a VEV the $\ov\tau$ and
$\ov\mu$ modes pair up.
After bringing down two copies of this terms and integrating out the fermionic zero modes, one is left with the measure
\bea
\int d{\cal M}_{I} &=&\int d^4x \,d^2 \theta \, \prod_a d\lambda_a \prod_b d\ov\lambda_b \int \,   d|m| \,
|m|^3 \, \delta\left(|m|^2 - {\xi}/{2}\right).
\eea
This is encouraging as with the $\ov\tau$-modes dropping
out everything seems to point  towards a superpotential contribution.
It only remains to absorb the matter zero modes $\lambda_a$, $\ov\lambda_b$
 which
were forced upon us by $U(1)_E$ invariance of the zero mode measure.
Recall that the sum of all charges of these fields is
$Q_E=\sum_a Q_E(\lambda_a)+\sum_b Q_E(\ov\lambda_b)=4$.
It is clear that pairs of such zero modes with opposite
$U(1)_E$ charge can generate the usual matter field couplings of the
type
\bea
             \lambda_a\,   \phi_{ab} \,  \ov\lambda_b,
\eea
but there will always be the surplus of four zero modes of type
$\ov\lambda_b$.

\begin{figure}[h]
\begin{center}
 \includegraphics[width=0.25\textwidth]{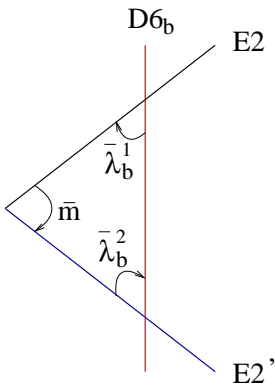}
\end{center}
\caption{\small Absorption of $\ov\lambda_b$ zero modes.}\label{fermifig}
\end{figure}

As shown in figure \ref{fermifig}\footnote{Figure \ref{fermifig}
displays the case with no additional matter field, namely
$<\bar{m}\,\bar{\lambda} \,\bar{\lambda}>$.}, due to the $U(1)_E$
charge the only way to absorb these extra $\ov\lambda$ zero modes is
via couplings of the type \bea \label{coupli}
           \ov m^{-1} \,  {\ov\lambda'}_b^{-1/2}\, \prod \phi_{b_i c_i}^{1}\,
             \ov\lambda_c^{-1/2}\,
\eea always involving the field $\ov m$. In (\ref{coupli}) the upper
index indicates the world-sheet charge $Q_{ws}$. Since all the
fields except $\ov m$ are chiral (in the sense of the ${\cal N}=2$
world-sheet supersymmetry) and $\ov m$ itself is anti-chiral, the
chiral ring structure tells us that all couplings of type
(\ref{coupli}) are {\it vanishing}: When we apply the picture changing operator to $ \ov
m^{-1}$ we do not pick up the right pole structure for a non-zero
amplitude \cite{Distler:1988ms}. On the other hand, with no additional matter
field $\phi$ in \eqref{coupli}, the amplitude is vanishing right away due to
violation of the $U(1)$ world-sheet charge.

Therefore, we conclude that in contrast to naive expectations, the
recombined $E2'-E2$ instanton cannot contribute to the
superpotential. There always remain four charged fermionic zero modes
which cannot be absorbed in a chiral manner. \\

For {\bf case II} there are no bosonic zero modes from the $E2'-E2$ intersection
and  therefore no brane recombination.
One only has the fermionic zero modes $\mu^{\alpha}$ of $U(1)_E$
charge $Q_E=2$. In this case we can write down the four-fermion
coupling
\bea
\label{coupli4f}
           \theta_{\alpha}^{3/2} \, (\mu^{\alpha})^{-1/2} \,
       {\lambda'}_a^{-1/2}\,  \lambda_a^{-1/2}\, ,
\eea
where again upper indices denote the $U(1)_{ws}$ charges.
Therefore, two such couplings can absorb the eight
appearing zero modes $\theta$, $\mu$, $\lambda^i_a$, $(\lambda'_a)^i$  so that one is only left
with the measure
\bea
\label{measureII}
\int d{\cal M}_{II} &=&\int d^4x \,d^2 \ov\tau \,
 \prod_a d\lambda_a \prod_b d\ov\lambda_b    \, ,
\eea where the total $U(1)_E$ charge of all the matter zero modes
$\lambda_a$ and $\ov\lambda_b$ vanishes. There is no way to absorb
the remaining $\ov\tau$ modes involving open string operators:
Clearly no superpotential terms are generated, as couplings like
\bea \label{coupli4fa}
           (\ov\tau_{\dot \alpha})^{-3/2} \, \
       {\ov\lambda}_a^{-1/2}\,  (\psi^{\alpha}_{ab})^{-1/2}\, \lambda_b^{-1/2}\,
\eea
are not allowed by Lorentz invariance and non-holomorphic interactions of the form
\bea
\label{coupli4fb}
           (\ov\tau_{\dot \alpha})^{-3/2} \, \
       {\ov\lambda}_a^{-1/2}\,  (\ov \psi^{\dot \alpha}_{ab})^{1/2}\, \lambda_b^{-1/2}\,
\eea
vanish as a consequence of $U(1)$ worldsheet  charge violation. By contrast, it is possible to absorb the $\ov \tau$- modes through couplings to anti-chiral fermions in the closed string sector of the form $\langle\ov \tau^{\dot \alpha} \ov \chi_{\dot \alpha} \rangle$, which will be discussed in section \ref{Kahler}. Clearly, the induced interactions are non-holomorphic and thus non-supersymmetric. This is however no wonder since, as we recall, the very presence of the effective $O^+$-plane leading to this kind of orientifold projection does not admit supersymmetric tadpole cancellation.

We conclude that in contrast to expectations based on spacetime-filling brane recombination processes, instanton recombination
does not lead to new $O(1)$ instantons which can contribute
to the superpotential. The reason is that due to
$U(1)_E$ charge conservation and the tadpole cancellation
conditions there arises a net number of  charged fermionic
matter zero modes which cannot be absorbed by chiral couplings.

For the Type IIB dual orientifold models
this observation implies that magnetised
$E5-E5'$ recombination, i.e. instantons carrying extensions (\ref{extens}) of
line bundles, do not generate
superpotential contributions either. The only known contributions in this
case come from $E1$-instantons wrapping  holomorphic
curves on the mirror Calabi-Yau manifold.

\subsection{Recombination of non-chiral  $E2-E2'$ instantons }

The deeper reason why chiral $E2-E2'$ intersecting instantons as in
case I do not lead, after brane recombination, to $O(1)$ instantons
seems to be that this $E2-E2'$ system carries charge along the
``directions'' of the orientifold $O6^-$ plane. In the Type IIB dual
situation this means  that the magnetised $E5-E5'$ system carries
$E5$-brane charge.

Consequently, it may be more promising to start with a magnetised
$E3-\ov{E3}'$ system which after brane-recombination only contains
$E1$-charge. Such a system necessarily has $E3\circ \ov E3'=0$ and
can only support  vector-like zero modes on the intersection. This
immediately implies that there are no $U(1)_E$ charged matter zero
modes necessary to ensure $U(1)_E$ invariance of the zero mode
measure.

The simplest non-trivial case involves one vector-like pair
of zero modes, i.e.
\bea
\label{int_pattern4}
[\Xi' \cap \Xi]^+=[\Xi' \cap \Xi]^- = 1, \ \
 [ \Pi_{O6} \cap \Xi]^+ =[ \Pi_{O6} \cap \Xi]^-= 1.
\eea
Therefore, for an $O6^-$-plane we have the zero modes
shown in figure \ref{antizeronc}.

\begin{table}[h]
\centering
\begin{tabular}{|c|c|}
\hline
 zero mode & $(Q_E)_{Q_{ws}}$  \\
\hline \hline
 $m, \ov m$ & $(2)_1$ ,$(-2)_{-1}$ \\
 $\ov\rho^{\dot \alpha}$ & $(-2)_{1/2}$ \\
\hline
 $n, \ov n$ & $(-2)_1$ ,$(2)_{-1}$ \\
 $\ov\nu^{\dot \alpha}$ & $(2)_{1/2}$ \\
\hline
\end{tabular}
\caption{Charged zero modes on non-chiral $E2-E2'$ intersection with
$O6^-$ plane% \vspace{3mm}
\label{antizeronc} }
\end{table}
There is still the fermionic coupling
\bea
\label{rec-actionbnc}
S_{E2}=  \ov \tau_{\dot \alpha}\left( m \,\ov \rho^{\dot \alpha} -
                n \,\ov \nu^{\dot \alpha} \right)
\eea
so that the $\ov\tau^{\dot\alpha}$ modes absorb one linear
combination of the fermionic zero modes.
% $\ov \mu^{\dot \alpha}$ of the $\ov\rho^{\dot\alpha}, \ov\nu^{\dot\alpha}$ modes given by
%\bea
%\ov \mu^{\dot \alpha} =  m \,\ov \rho^{\dot \alpha} +
%                n \,\ov \nu^{\dot \alpha},
%\eea
%while the orthogonal mode
%\bea
%\ov {\widetilde \mu}^{\dot \alpha}  =  m \,\ov \rho^{\dot \alpha} -
%                n \,\ov \nu^{\dot \alpha}
%\eea
%remains in the measure.
In addition the single real bosonic D-term constraint\footnote{One
might expect that similar to the ADHM construction of gauge
instantons one has three D-term constraints. But from the $U(1)_E$
and $U(1)_{ws}$ charges in Table \ref{antizeronc} it is clear that
one can build only the neutral combination in eq. (\ref{dtermnc}).}
\bea \label{dtermnc}
           m^2_{1}\, \ov m^{-2}_{-1} - n^{-2}_{1}\, \ov n^{2}_{-1}=0
\eea fixes \bea \label{absD} m \, \ov m  = n \, \ov n, \eea
%upon writing
%\bea
%m= |m| \, e^{i \alpha}, \quad\quad n= |n| \, e^{i \beta}.
%\eea
where the lower index denotes the $U(1)_{ws}$ charge while the upper one
refers to $U(1)_E$. For initially rigid instantons, i.e. in the
absence of $E2$-reparametrisation moduli, there exist no F-term
constraints which would prevent a non-vanishing VEV  $m \, \ov m  =
n \, \ov n\ne 0$ corresponding to brane recombination.

As in the analogous process for chiral intersections, recombination
breaks the $U(1)_E$. The associated gauge degree of freedom can be
used to set \bea m= n \eea as opposed to merely (\ref{absD}).
Integrating out the $\ov\tau$ modes together with the linear
combination $\ov \mu=\ov\rho - \ov \nu$ of fermionic zero modes as
appearing in (\ref{rec-actionbnc}) brings down a factor of $m^2$.
%removes one of the two phases $\alpha$, $\beta$  from the measure in that it is absorbed by the Higgsing of the U(1) gauge group acting on the instanton.

After recombination,  one is left  with the measure
%\footnote{This uses that $d^2 \ov \rho^{\dot \alpha} d^2 \, \ov \nu^{\dot \alpha} = 4 (2m)^2 \, (2n)^2 \, d^2 \ov {\mu}^{\dot \alpha} \,  d^2 \ov {\widetilde \mu}^{\dot \alpha} $ (as follows from $\ov \rho^{\dot \alpha} = (2m)^{-1}( \ov {           \mu}^{\dot \alpha} +  \ov {\widetilde \mu}^{\dot \alpha}, \ov \nu^{\dot \alpha} = (2n)^{-1}( \ov { \mu}^{\dot \alpha} -  \ov {\widetilde \mu}^{\dot \alpha}  $)  as well as $ d m_1\,  d\ov{m}_{-1}=  |m|\, d|m| \, d \alpha  $. }
\bea
\label{dreimeasure}
\int d{\cal M}_{III} &=&\int d^4x \,d^2 \theta \,  d^2 \ov {\widetilde \mu}^{\dot \alpha}_{1/2}
     \, d m_1\,  d\ov{m}_{-1}\, m_1^2,
%\int d{\cal M}_{III} &=&\int d^4x \,d^2 \ov\theta \, d^2 \ov {\widetilde \mu}^{\dot \alpha}_{1/2} \, d \beta \, \, |m|^6 e^{6 i \beta},
\eea where again the lower index denotes the $U(1)_{ws}$ charge in
the canonical ghost picture and $\ov {\widetilde \mu}^{\dot
\alpha}_{1/2}= \ov\rho + \ov \nu$  stands for the remaining linear
combination of fermionic zero modes. In addition, there can of
course be charged zero modes $\lambda_a$, $\ov \lambda_{b}$.

Ignoring the additional factor of $ m_1^2$ for the moment, this zero mode structure is precisely that of an $O(1)$ instanton
with one deformation $b_1(\Xi)=1$ of the first type (see discussion around (\ref{measure_family}))).
From our discussion in section \ref{sec_BW} we expect this configuration to generate higher fermionic F-terms of Beasley-Witten type.

%In the dual Type I model
%it is  identical to the zero mode structure
%of $E1$-instantons moving in a one-dimensional family as discussed
%in \cite{Beasley:2005iu}.

%Beasley-Witten found that such instantons can generate
%higher fermionic couplings for the closed string
%fields \cite{Beasley:2005iu}. In our case, there exist, in addition to these closed string couplings, terms which
%generate higher fermi-couplings also for the matter fields.

Extrapolating from the CFT of the $E2-E2'$ sector before recombination, the relevant couplings after recombination are inherited from
\bea
\label{BWcoup1}
        (\ov{m}_{-1} \ov\nu_{1/2}^{\dot\alpha} + \ov{n}_{-1} \ov\rho_{1/2}^{\dot\alpha}) \,\, \lambda^a_{-1/2}\,\,
            (\ov\psi_{1/2})_{\dot\alpha}\,\, \ov \lambda^b_{-1/2} \longrightarrow \ov{m}_{-1} \ov {\widetilde \mu}^{\dot \alpha}_{1/2}  \lambda^a_{-1/2}\,\,
            (\ov\psi_{1/2})_{\dot\alpha}\,\, \ov \lambda^b_{-1/2},
\eea
where the fermionic matter field $\ov\psi_{1/2}^{\dot\alpha}$
lives at the intersection $D6_a-D6_b$ and lies in the anti-chiral
superfield $\ov\Phi=\ov \phi + \ov\tau \ov\psi$.
Note that the above coupling does not violate any of the general ${\cal N}=2$ SCFT selection rules so that even without a direct computation we expect it to be present for sufficiently generic backgrounds.
Integrating out two copies of this interaction term brings down the fermion bilinear $\ov\psi_{1/2}\ov\psi_{1/2} $ characteristic for the higher fermionic terms described in \cite{Beasley:2005iu} as well as a factor of  $\ov m_{-1}^2$. The bosonic measure can then be brought into standard form by a simple change of variables with $\widetilde m = m^3$ and we are left with
\bea
\int d{\cal M}_{III} &=&\int d^4x \,d^2 \theta \, d\widetilde m_1 \, d\ov{\widetilde m}_{-1}\, \,\ov\psi_{1/2}\ov\psi_{1/2} .
\eea
Together with the chiral fermion bilinear pulled by the two $\theta^{\alpha}$ modes  this results in the four-fermi terms as discussed in section \ref{sec_BW}.

Its bosonic derivative superpartner involves absorbing one pair of $\theta \ov \mu$ in a coupling of the form (after recombination)
\bea
\label{BWcoup2}
\ov{m}_{-1} \,  (\ov \tau_{\dot \alpha})_{3/2} \,\, \ov {\widetilde \mu}^{\dot \alpha}_{1/2} \,\,  \lambda^a_{-1/2}\,\, \ov \phi_{-1} \,\, \ov \lambda^b_{-1/2}.
\eea
With $\ov \phi_{-1}$ and $\ov{m}_{-1}$ in the zero ghost picture\footnote{Note that for $\ov{m}_{-1}$  the PCO can only act non-trivially in the internal part since its vertex does not carry any momentum.} this  generates a derivative for the boson $ \ov \phi_{-1}$. Bringing down two copies of this term indeed yields the derivative superpartner to the above four-fermi term, again in agreement with \cite{Beasley:2005iu}.

\subsection{Contribution to superpotential}
\label{sec_super?}

\begin{figure}[ht]
\begin{center}
 \includegraphics[width=0.4\textwidth]{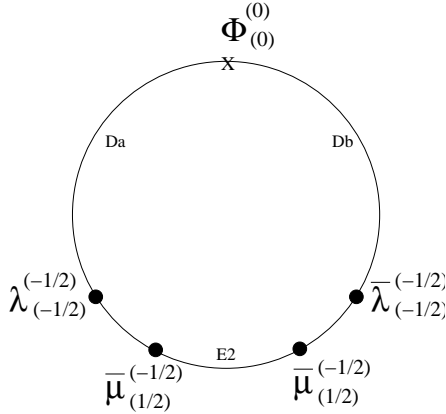}
\end{center}
\caption{\small Absorption of $\ov{\mu}^{\dot\alpha}_{1/2}$
   for superpotential contributions. The upper indices are the
ghost sectors and the lower ones the $Q_{ws}$ charges.}\label{superpotfig}
\end{figure}

It has been observed for world-sheet instantons in the
heterotic string that instantons moving in a family not only generate higher fermionic F-terms, but
can also contribute to the superpotential \cite{Dine:1987bq}.
Recall that such instantons are dual to $E2$-instantons
with deformations of the first kind and with
a zero mode structure as in (\ref{measure_family}) for each deformation.
As we just saw, recombination of a non-chiral $E2$-$E2'$ pair yields precisely such objects.
For superpotential contributions to exist it must be
possible to absorb the fermionic zero modes
without generating higher fermionic or derivative terms as in (\ref{BWcoup1}) or (\ref{BWcoup2}).
A way to do this for matter field superpotential contributions
is shown in figure \ref{superpotfig}. There $\ov \mu$ denotes the fermionic reparameterisation mode independently of whether the instanton is the result of recombination or not. In the first case, we should actually replace $\ov \mu$ by $\ov{m}_{-1} \ov {\widetilde \mu}$ as before.

\noindent
If this five point function has a contact term
and if the remaining integral over the
bosonic instanton moduli space does not vanish,
then a contribution to the superpotential can
be generated. We stress again that from a general ${\cal N}=2$ SCFT point of view, no obvious selection rules forbid such an interaction term. Having said this, one can easily convince oneself that for factorizable three-cycles on toroidal orbifolds the amplitude vanishes due to violation of the $U(1)$ worldsheet charge which has to be conserved for each of the three tori separately. This, however, need not be so for more general setups.

\noindent By contrast, it is clear that these disc amplitudes vanish for
$E2$-deformations of the second kind as defined in \cite{Akerblom:2007uc}. Recall from section \ref{sec_BW}
 that these give rise to chiral instead of anti-chiral deformation modulini.

\noindent To summarize, non-chiral $E2-E2'$ recombination results in
an object with at least two bosonic and two fermionic
zero modes from a surviving deformation of the first kind
of the recombined instanton. These objects can
generate higher fermion couplings and under certain circumstances
can also  contribute to the superpotential.

\section{F-term correction to complex structure moduli space}
\label{Kahler}

%Before instanton recombination a complex $U(1)$ instanton
%can contribute to the complex structure moduli
%K\"ahler potential.

Having analysed the consequences of  zero modes in the $E2-E2'$
sector in addition to the four Goldstinos for a $U(1)$ instanton, in
this section we are interested in the induced couplings if the
uncharged measure merely takes the form \bea
\label{unchargedmeasure_a} \int d^4 x\, d^2\theta\, d^2 \ov\tau \eea
in the first place. Consider therefore a rigid $U(1)$ instanton with
the geometric intersection numbers \bea \label{int_pattern3} \Xi'
\cap \Xi = 0 =   \Pi_{O6} \cap \Xi. \eea This is easily realised
e.g. for cycles parallel to, but not on top of the orientiold plane
in some subspace. The uncharged zero mode measure
(\ref{unchargedmeasure_a}) is to be supplemented by additional
charged zero modes $\lambda$ if present. Since there are no zero
modes in the  $E2-E2$' sector which would be sensitive to the
orientifold action, we might expect this type of instantons to be
describable in terms of half-BPS instantons of the underlying ${\cal
N}=2$ supersymmetry preserved by the internal Calabi-Yau before
orientifolding. The correction to the complex structure moduli space
metric by $E2$-instantons in type IIA Calabi-Yau compactification
has been discussed recently in
\cite{Halmagyi:2007wi}\footnote{Recall that the local factorisation
of the moduli space describing the vector and hypermultiplets in
general  ${\cal N}=2$ compactifications \cite{deWit:1984px} forbids
corrections to the K\"ahler moduli since the dilaton sits in a
hypermulitplet.}. Following this logic, we would anticipate the
generation of $E2$-corrections to the complex structure K\"ahler
potential by the $U(1)$ instanton described by
(\ref{unchargedmeasure_a}).
%Let us first ignore the possibility of charged zero modes $\lambda$ and focus entirely on non-perturbative couplings in the closed string sector.
%An important guideline is the local factorisation of the moduli space describing the vector and hypermultiplets in general  ${\cal N}=2$ compactifications \cite{deWit:1984px}. Since the dilaton sits in a hypermultiplet and the instanton induced couplings depend exponentially on $1/g_s$, D-brane instantons in type II compactifications can in general correct only the hypermultiplet couplings. For the case of $E2$-instantons in type IIA this means that only the complex structure moduli space can be corrected by the instanton \cite{Halmagyi:2007wi}. Clearly, the instanton measure  (\ref{unchargedmeasure_a}) for rigid uncharged $E2$-instantons contains the four fermionic zero modes required for corrections of the K\"ahler potential.

However, while the chiral Goldstino modes $\theta$ are indeed associated with the breakdown of the ${\cal N}=1$ subalgebra of this ${\cal N}=2$ symmetry which is preserved by the orientifold, their anti-chiral partners  $\ov \tau$ correspond to the orthogonal ${\cal N}=1$ subalgebra. The above measure (\ref{unchargedmeasure_a}) does therefore not cover the full ${\cal N}=1$ superspace as required for the generation of a K\"ahler potential. Rather, the integral is only over half of the ${\cal N}=1$ superspace. While this calls for the generation of an F-term as opposed to a D-term, the additional fermionic zero modes $\ov \tau$ will  result in higher fermonic couplings of Beasley-Witten type discussed in detail in section \ref{sec_BW}.

An important difference to the F-terms discussed previously is that
now only the complex structure moduli  receive derivative
corrections. Denote by $w$ and $a$ the scalar and axionic parts of
the scalar component $U= w - i { a}$ of a complex structure
superfield. Then evaluation of the amplitudes $\langle \theta \, \ov
w \, \ov\tau \rangle$ and $\langle \theta \,\ov a \, \ov\tau
\rangle$ gives rise to the terms \bea \theta \,\sigma^{\mu}\,
\bar{\tau} \,\, \partial_{\mu}\, \ov w^{i}, \quad\quad \theta
\,\sigma^{\mu}\, \bar{\tau} \,\, \partial_{\mu}\, \ov a^{i} \eea in
the moduli action. For the details of this computation in the
context of toroidal orbifolds see appendix \ref{closed_int}. The
absence of analogous terms for the K\"ahler moduli is a consequence
of $U(1)$ worldsheet charge conservation. Integrating out two copies
thereof indeed generates a derivative coupling of the form $
e^{-S_{E2}} \, \partial \ov U \partial \ov U$. Together with their
fermionic partners, the derivative F-terms can be summarized by \bea
\label{BW_3} S = \int d^4 x \, d^2 \theta \, \, e^{-{\cal U}(\Xi)}
\,\, f_{\ov i, \ov j}\left(e^{{\cal T}_i}, e^{\Delta_i} \right)\,\,
{\ov{\cal D}}^{\dot \alpha} {\ov {\cal U}}^{\ov i} \,  {\ov{\cal
D}}_{\dot \alpha} {\ov {\cal U}}^{\ov j} + \, \, \, h.c., \eea where
the complex conjugate part is due to the anti-instanton
contribution. Note the difference to eq. (\ref{BW_2}) describing the
higher fermionic terms for $E2$-instantons with deformation modes.
In the presence of charged zero modes $\lambda$ these F-term
corrections for the complex structure moduli involve appropriate
powers of charged open string fields required to soak up the
$\lambda$ modes. This amounts to replacing $e^{-{\cal U}(\Xi)}
\rightarrow e^{-{\cal U}(\Xi)} \prod_i \Phi_{a_i,b_i}$.

\section{Flux-induced lifting of zero modes}
\label{sec_fluxes}

The additional two zero modes $\ov \tau^{\dot \alpha}$ which, if present, prevent the generation of a superpotential by the instanton,
are a consequence of the underlying ${\cal N}=2$ supersymmetry preserved in the bulk of the Calabi-Yau away from the orientifold plane in the way described in section \ref{sec_supo}. It has therefore been speculated in the literature \cite{Argurio:2007vq,Ibanez:2007rs,Aharony:2007pr}  that these Goldstinos might be lifted in the presence of suitable background fluxes. An intuitive reason why this could be the case is that under appropriate circumstances the instanton is expected to feel only the ${\cal N}=1$ supersymmetry preserved by the  flux in the bulk. In such situations the $\ov \tau$ modes are not protected as the Goldstinos of the orthogonal ${\cal N}=1$ supersymmetry and it might be possible that indeed only the two $\theta^{\alpha}$ modes remain massless in the universal zero modes sector.

While our previous presentation has focused on D-brane instantons in Type IIA orientifolds, the natural arena to study the effects of background fluxes is the framework of Type IIB compactifications, where we can take advantage of the by now quite mature understanding  of a fully consistent incorporation of supersymmetric three-form flux  (for references see e.g. \cite{Grana:2005jc,Douglas:2006es}). The lifting of fermionic zero modes by supersymmetric three-form flux has been analysed in special cases in \cite{Tripathy:2005hv,Bergshoeff:2005yp,Park:2005hj,Lust:2005cu} in the context of $E3$-instantons wrapping a holomorphic divisor of the internal (conformal) Calabi-Yau. The most general such situation involves the presence also of supersymmetric gauge flux on the worldvolume $E3$-brane. This corresponds on the Type IIA side to $E2$-instantons at general angles with the $O6$-plane and is the configuration we are primarily interested in. To the best of our knowledge the possible consequences of such gauge fluxes on the zero mode structure have not been analysed explicitly so far.
Before addressing the more general case, we review first the situation of vanishing gauge flux.

\subsection{Zero mode lifting for unmagnetised E3-instantons}

In the spirit of  \cite{Giddings:2001yu}, we consider Type IIB
orientifold compactifications  with  an ${\cal N}=1$ supersymmetric
combination $ G = F - \tau H $ of RR and NS flux $ F= d C_{2}$ and $
H= dB$ such that the complexified dilaton $\tau = C_0 + i e^{-\phi}$
is  constant. The internal manifold is therefore conformally
Calabi-Yau with constant warp factor. In order to  preserve
supersymmetry, the flux has to be of (2,1) type\footnote{In the
presence of a non-perturbative superpotential this condition is
relaxed to include also (0,3) components \cite{Lust:2005cu}.} and
satisfy the primitivity condition $J \wedge G =0$ in terms of the
K\"ahler form $J$. We consider an $E3$-brane wrapping a holomorphic
divisor $\Gamma$. Since our interest here focuses on the lifting of
$\ov \tau$-modes, we assume that $\Gamma$ is not invariant under the
holomorphic involution $\sigma$ defining the orientifold action
$\Omega (-1)^{F_L} \sigma$ so that the $\ov \tau$-modes are not projected out. For the simple setup of unmagnetised divisors, we can then simply identify the instanton with its orientifold image and focus on the instanton action before orientifolding without further ado.

The part in the $E3$-brane worldvolume action describing the coupling of such three-form flux to the (uncharged) zero modes  $\omega$\footnote{The corresponding objects in \cite{Bergshoeff:2005yp} are called $\theta$, see eq. (4.1.). Recall that we reserve the notation $\theta$ and $\ov \tau$ for the four-dimensional spinor associated with the universal zero modes.} reads \cite{Marolf:2003vf,Bergshoeff:2005yp}
\bea
\label{Fluxaction1}
S= \int_{\Gamma} d^4 \zeta\sqrt{ {\rm det}g}\, \, \omega \, \,  \left(e^{-\phi} \, \Gamma^{\tilde m} \nabla_{\tilde m} + \frac{1}{8} \, \widetilde G_{\tilde m \tilde n p} \, \Gamma^{\tilde m \tilde n p} \right) \omega.
\eea
The combination $\widetilde G_{\tilde m \tilde n p}$ appearing above is defined  as
$\widetilde G_{\tilde m \tilde n p}= e^{-\phi} H_{\tilde m \tilde n p} + i F_{\tilde m \tilde n p}' \gamma_5$ in terms of  $F_{\tilde m \tilde n p}'= F_{\tilde m \tilde n p} - C_0 H_{\tilde m \tilde n p}$ and the four-dimensional matrix $\gamma_5$. The indices $\tilde m,\tilde n$ are along the four-cycle $\Gamma$ and $p$ is transverse to it. While the above action was derived in \cite{Marolf:2003vf,Bergshoeff:2005yp} entirely with the help of supergravity methods, one could in principle determine it by analysing the CFT coupling of the closed string fields to the boundary, see \cite{Green:1997tv,Lust:2004cx,Bertolini:2005qh,Billo:2006jm} for the relevant techniques.

The Euclidean action (\ref{Fluxaction1}) uses a particular gauge fixing condition to eliminate the unphysical degrees of freedom due to $\kappa$-symmetry (cf. eq. 4.9 of \cite{Bergshoeff:2005yp}). As a result, the spinor $\omega$ is a sixteen-component Weyl spinor since we consider a Euclidean action.
Locally, we can choose complex coordinates $a,b = 1,2$ along $\Gamma$ and $z,\ov z$ for the transverse direction.
It is convenient to use the standard definition of the Clifford vacuum
$|\Omega\rangle$,
\bea
\Gamma^z |\Omega\rangle =0, \quad \Gamma^a |\Omega\rangle =0
\eea
and to decompose the spinor $\omega$ into its external and internal part.
The latter can be grouped according to its chirality along the normal bundle of the divisor as
\bea
\label{spinor_decomp}
\epsilon_+ &=& \phi |\Omega\rangle +  \phi_{\ov a}\Gamma^{\ov a} |\Omega\rangle + \phi_{\ov a \ov b}\Gamma^{\ov a \ov b} |\Omega\rangle, \nonumber \\
\epsilon_- &=& \phi_{\ov z} \Gamma^{\ov z}|\Omega\rangle +  \phi_{\ov a\ov z}\Gamma^{\ov a\ov z} |\Omega\rangle + \phi_{\ov a \ov b\ov z}\Gamma^{\ov a \ov b\ov z} |\Omega\rangle.
\eea
In this language we can immediately identify the universal fermionic zero modes with four-dimensional polarisation $\theta^{\alpha}$ and  ${\overline \tau}^{\dot\alpha}$ as given by
\bea
\label{universal_decomp}
\omega_0^{(1)} = \theta \otimes  \phi |\Omega\rangle, \quad\quad\quad  \omega_0^{(2)} = \overline \tau \otimes  \phi_{\ov a \ov b\ov z}\Gamma^{\ov a \ov b\ov z} |\Omega\rangle.
\eea
The fact that they are the "universal" zero modes follows from their correspondence with the cohomology group $H^{(0,0)}(\Gamma)$.
The remaining components in (\ref{spinor_decomp}) are associated with the reparametrisation modulini and Wilson line fermions of the four-cycle counted by $H^{(0,2)}(\Gamma)$ and $H^{(0,1)}(\Gamma)$, respectively \cite{Jockers:2004yj,Bergshoeff:2005yp,Lust:2006zg}.

Starting from the above action, i.e. in the absence of gauge flux, \cite{Bergshoeff:2005yp} computed the remaining zero modes in the presence of primitive (2,1) three-form flux. In particular, their analysis shows that the four universal zero modes (\ref{universal_decomp}) are not lifted in such a situation. In fact, one can easily convince oneself that the zero mode $\omega_0^{(2)}$ does not couple to primitive (2,1) flux.
E.g. %\footnote{The remaining cases ${\widetilde G}_{a \ov z z}$ and ${\widetilde G}_{a b \ov z}$ are treated similarly.}
\bea
{\widetilde G}_{\ov a b z} \Gamma^{\ov a b z} \Gamma^{\ov 1}  \Gamma^{\ov 2}  \Gamma^{\ov 3 }|\Omega\rangle &=& {\widetilde G}_{\ov a b z} g^{b \ov1} g^{z \ov 3} \Gamma^{\ov a}  \Gamma^{\ov 2}|\Omega\rangle  - {\widetilde G}_{\ov a b z} g^{b \ov2} g^{z \ov 3} \Gamma^{\ov a}  \Gamma^{\ov 1}|\Omega\rangle \nonumber \\
&=&  {\widetilde G}_{\ov 1 b z} g^{b \ov1} g^{z \ov 3} \Gamma^{\ov 1}  \Gamma^{\ov 2}|\Omega\rangle  + {\widetilde G}_{\ov 2 b z} g^{b \ov2} g^{z \ov 3} \Gamma^{\ov 1}  \Gamma^{\ov 2}|\Omega\rangle \nonumber \\
&=& {\widetilde G}_{\ov a b z}  g^{b \ov a}  g^{z \ov 3} \Gamma^{\ov 1}  \Gamma^{\ov 2}|\Omega\rangle  = 0.
\eea
The last equation follows from the identity \cite{Bergshoeff:2005yp}
\bea
{\widetilde G} |\Omega\rangle = i \, {G} |\Omega\rangle
\eea
together with primitivity of $G$,
\bea
g^{c \ov c'} G_{b c \ov c'} = 0.
\eea

Likewise, potential $(0,3)$ components of G-flux can be shown not to couple to the universal modes. This type of flux is allowed by the equations of motion and supersymmetric once the non-perturbative superpotential is taken into account in the analysis of the gravitino variation \cite{Curio:2005ew,Lust:2005cu}.

\subsection{Zero mode lifting for magnetised E3-instantons}

We are now ready to address our main question, the inclusion of non-trivial gauge flux on the instanton.
The worldvolume action of the $E3$-instanton contains, in addition to (\ref{Fluxaction1}), two pieces linear and quadratic in the gauge invariant combination  ${\cal F} = F_{\rm gauge} - B$ of the worldvolume gauge field and Neveu-Schwarz two-form.
Since we are considering an orientifold, we have to add the contribution of the $E3$-instanton together with its image under $\Omega (-1)^{F_L} \sigma$. As described in \cite{Jockers:2004yj}, this amounts to considering the instanton wrapping the divisor $\widetilde \Gamma= \Gamma + \sigma \Gamma$ and to expand the worldvolume fields, according to their parity under $\sigma$, into their components along the invariant and anti-invariant cohomology on $\widetilde \Gamma$.
Since ${\cal F}$ is anti-invariant under $\Omega$, the linear terms in the action survive only for the components of ${\cal F}$ along elements of $H_{(1,1)}^-(\widetilde \Gamma)$.
%, while those quadratic in ${\cal F}$ survive along elements of $H_{(1,1)}^+(\tilde \Gamma)$.

Before orientifolding, the relevant part of the quadratic term is the sum of the two terms\footnote{Note that for simplicity, we are using here the gauge of \cite{Tripathy:2005hv}, eq. (29). This $\kappa-$symmetry fixing is different from the one in which (\ref{Fluxaction1}) is written and corresponds essentially to the one  of \cite{Aganagic:1996pe,Bergshoeff:1997kr}. As emphasized in \cite{Bergshoeff:2005yp,Bandos:2006wb} the gauge fixing condition and the orientifold projection have to be compatible for branes invariant under the orientifold. Since we are interested in the more general situation of non-invariant branes or instantons, it suffices for our purposes to work in the gauge of \cite{Tripathy:2005hv}.} \cite{Tripathy:2005hv}
%\footnote{Note that the result (\ref{Fluxaction1}) of \cite{Bergshoeff:2005yp} and t he action derived in \cite{Tripathy:2005hv} are formulated using different gauge fixing conditions for the $\kappa$-symmetry. }
\bea
\label{action_gaugeflux}
S_{\rm DBI} &=& %- \frac{\mu}{48}
- \frac{ \mu}{48} \int_{\Gamma} d^4 \zeta\sqrt{ {\rm det}g}\, \, \omega \, \Gamma^{\tilde m \tilde n p} \,\, \omega \, \, e^{-\phi} H_{\tilde m \tilde n p}  \, \left( \frac{1}{4} \,\,{\cal F}^2 \right), \nonumber \\
S_{\rm WZ} &=& - \frac{\mu}{48}
\int_{\Gamma} \, \, \omega \,  \Gamma^{\tilde m \tilde n p}  \,\, \omega \,\, ( i F'_{\tilde m \tilde n p}) \, \left( \frac{1}{2} \, \, {\cal F} \wedge {\cal F} \right).
\eea
In four Euclidean dimensions, solutions to the field equations and Bianchi identity can be taken to satisfy the self-duality constraint ${\cal F}= \star {\cal F}$. Together with $\int \sqrt{ {\rm det}g}  \left( \frac{1}{4} \,\,{\cal F}^2 \right) = \int \frac{1}{2} {\cal F} \wedge \star {\cal F}$
we find that the relevant couplings combine into
\bea
- \frac{\mu}{48} \int_{\Gamma} d^4 \zeta\sqrt{ {\rm det}g}\, \, \omega \,  \, G_{\tilde m \tilde n p} \, \Gamma^{\tilde m \tilde n p} \,  \omega\,\,\left(\frac{1}{4} \,\,{\cal F}^2 \right).
\eea
By the same reasoning as above, this interaction does not induce any mass terms for the universal zero modes provided we stick to supersymmetric (2,1) (or even (0,3)) flux.

Let us now discuss if the term linear in ${\cal F}$ saves the day, given in the upstairs geometry by  \cite{Tripathy:2005hv}
\bea
\label{F_linear}
S &= &\frac{\mu}{16}
\int_{\Gamma} d^4 \zeta\sqrt{ {\rm det}g} \left( {\cal F}_{\tilde i \tilde k} \,\, \omega \, \Gamma ^{{\tilde k} st} \,  e^{- \phi} \, {H^{{\tilde i}}}_{ st} \, \omega - \frac{i}{2} \epsilon^{\widetilde{i}\widetilde{j}\widetilde{k}\widetilde{l}}\, {\cal F}_{\widetilde{i}\widetilde{j}} \,\, \omega {\Gamma_{\widetilde k}} ^{st}\, (F')_{{\widetilde l} st} \, \omega \right) \nonumber \\
&=& -i \,\frac{\mu}{16}  \,\int_{\Gamma} d^4 \zeta\sqrt{ {\rm det}g} \, {\cal F}_{\widetilde i \widetilde j} \, \, \omega \Gamma^{\widetilde i st} \omega \,\, g^{ \widetilde j \widetilde k}\,\, G_{\widetilde kst}.
\eea
Again self-duality of the gauge flux, $\frac{1}{2} \epsilon^{\widetilde{i}\widetilde{j}\widetilde{k}\widetilde{l}}\, {\cal F}_{\widetilde{i}\widetilde{j}} = {\cal F}^{\widetilde{k}\widetilde{l}}$, is used and a tilde denotes indices parallel to the worldvolume, whereas $s,t$ are general internal indices.

While the index structure of the $\Gamma$-matrices is still of type $(2,1)$ due to contraction with the hermitian metric, the above action may in principle induce non-vanishing couplings involving the universal modes. After all, the vanishing of such couplings in the absence of gauge flux rested also upon primitivity of $G_3$, which is not necessarily satisfied by the combination of ${\cal F}$ and  $G_3$ contracted with the $\Gamma$ in (\ref{F_linear}).
As we stressed, these couplings, being linear in ${\cal F}$, only survive the orientifold action in the presence of anti-invariant two-cycles on the divisor ${\widetilde \Gamma}$. We will illustrate this issue in more detail in the next subsection.

On the other hand, in the absence of such cycles, as e.g. for the $T^6/{\mathbb Z}_2$ example studied in \cite{Kachru:2002he},
the $\ov \tau$-modes remain massless even after taking into account the backreaction of the three-form flux on the instanton moduli action. While this may seem counter-intuitive because  they are no longer protected as Goldstinos in the presence of three-form flux, this is just an example of the familiar fact even though all symmetries broken by the instanton result in associated zero modes, the converse need not be true.

\subsection{A simple example with linear gauge fields}

In the presence of suitable three-form flux and for non-vanishing gauge flux $F$, the linear term in ${\cal F}$ leads to
a coupling of the zero mode $\omega^{(2)}_0$
proportional to
\bea
\label{foehr}
{G}_{\ov a b z}  {F}^{b \ov a}  g^{z \ov 3} \Gamma^{\ov 1}  \Gamma^{\ov 2}|\Omega\rangle  .
\eea
As stated above, this does not vanish directly due the primitivity condition for $G$-flux
and the hermitian Yang-Mills equation for the gauge flux ${F}$.
Under the orientifold projection the flux components ${F} \in H^{+}_{1,1}(\widetilde \Gamma)$
are mapped to $-{F}$ and the associated terms in (\ref{foehr}) vanish trivially.
But for the components   $F \in H^{-}_{1,1}(\widetilde\Gamma)$  there is a chance that
the zero modes $\omega^{(0)}_2$ become massive.

More precisely, the action (\ref{F_linear}) leads to a coupling between $\omega^{(2)}_0$ and the mode $\phi_{\ov a \ov b}\Gamma^{\ov a \ov b} |\Omega\rangle$.
Integrating out the two types of zero modes lifts both the extra universal modes and the deformation modes $\phi_{\ov a \ov b}\Gamma^{\ov a \ov b} |\Omega\rangle$. For this mechanism to work, the deformations associated with $\phi_{\ov a \ov b}\Gamma^{\ov a \ov b} |\Omega\rangle$ have to be unobstructed, of course. On the other hand, the topological index $N_+ - N_{-}$ counting the difference between zero modes of positive and negative chirality with respect to the normal bundle of the divisor (see discussion around eq. \ref{spinor_decomp}) remains unchanged. This is reassuring  as by turning on suitable background B-field in addition to the gauge flux we may continuously set the quantity ${\cal F}$ appearing in the coupling (\ref{F_linear}) to zero, which should not change any topological quantities.

Let us illustrate this in a simple local example on a toroidal orientifold.
We compactify Type IIB on $T^6$ with metric
\bea
      ds^2= \sum_I dz_I\, d\ov z_I
\eea
and mod out by the orientifold projection $\Omega\sigma (-1)^{F_L}$
with $\sigma:z_2\to -z_2$.
Ignoring the resulting tadpole cancellation conditions for a second,
we now turn on $\Omega\sigma (-1)^{F_L}$ invariant $G_3$-form flux.
Consider  an $E3$-brane in this background
on the divisor $\Gamma$ given by the first two $T^2$s times a point on the third
one.
For vanishing
Wilson line along the first $T^2$, $\Gamma$ is invariant under
the orientifold projection and the instanton is of type $O(1)$. Since we are interested in lifting the ${\ov \tau}$-modes we assume the presence of a Wilson line rendering the instanton non-invariant.

On this $E3$-brane we turn on  constant
gauge flux of type
\bea
              { F}^{\ov 1 2} \in H^{-}_{1,1}(\widetilde \Gamma),
\eea
with $\widetilde \Gamma= \Gamma + \sigma \Gamma$ as before. This flux is  invariant under the orientifold projection and satisfies the HYM equation. Consistently, the
brane couples to the likewise invariant two-form $(C_2)_{\ov 1,2}$.

Then the coupling of the zero mode  $\omega^{(2)}_0$ on the
instanton is proportional to \bea
   {G}_{\ov a b z}  { F}^{\ov ab}={G}_{\ov 1 2 3}  { F}^{\ov
     1 2},
\eea which can be non-vanishing. Indeed the flux component ${G}_{\ov
1 2 3}$ is invariant under the orientifold projection. This simple
example shows that, ignoring tadpole constraints, it is possible
that the $\omega^{(2)}_0$ modes decouple for non-vanishing $G_3$
form flux.

However, when it comes to satisfying the tadpole constraints, we have to
introduce both further $D7$-branes to cancel
the $O7$-plane tadpole and  an $O3$-plane to cancel the tadpole induced
by the $G_3$-form.
The easiest way to get the $O3$-plane is to also mod out the model
by the ${\mathbb Z}_2$ action $z_{1,3}\to -z_{1,3}$, essentially turning the configuration into the fluxed $K3 \times \frac{T^2}{{\mathbb Z}_2}$ model studied in \cite{Tripathy:2002qw}.
However, in this case the $E3$  is not invariant under
this ${\mathbb Z}_2$, but mapped to an $E3$ brane with opposite
gauge flux $-{ F}^{\ov 1 2}$. Therefore, the  coupling
of the $\omega^{(2)}_0$ modes again trivially vanishes.
We leave it for future work to study more general concrete global models
of such a configuration in detail and to verify
if the $\ov \tau$-modes can actually be lifted.

\section{Conclusions}

This paper has investigated in detail under which circumstances D-brane instantons can contribute to the superpotential in Type II orientifolds.
A key role is played by the two universal zero modes $\ov \tau$ which are a remnant of the local ${\cal N}=2$ supersymmetry felt by instantons not invariant under the orientifold action. Their presence obstructs the generation of a superpotential.
If these modes are not lifted and in the absence of additional zero modes between the instanton and its orientifold image, the instanton generates higher-fermionic F-term corrections which in general depend also on open string operators. Previously, such terms had been considered in the context of heterotic worldsheet instantons moving in a family \cite{Beasley:2005iu}.

Our main interest has been in possible mechanisms to lift the $\ov \tau$ modes such that superpotential contributions are possible. Clearly, this question is of significance for an analysis of the quantum corrected
moduli space of string vacua as well as for determining the effective interactions in the vacuum.

We first focused on an  effect which, for $E2$-instantons in Type
IIA orientifolds, is describable as recombination of the instanton
with its orientifold image. Equivalently, we asked whether in the
Type IIB/Type I dual picture $E5$-instantons carrying non-trivial
extension bundles generate superpotential couplings. If so, this
would have important consequences  also for the heterotic string. We
found that while the $\ov \tau$-modes are indeed absent in such
situations, there arise generically additional  charged zero modes
which cannot be lifted, thus obstructing a contribution to the
effective action. By contrast, for the special case that the
instanton and its orientifold image preserve a common ${\cal N}=1$
supersymmetry, no such zero modes arise and the recombined object
can generate a superpotential provided its reparametrisation moduli
can be lifted. For general Calabi-Yau manifolds, we identified
appropriate open-string dependent couplings in the instanton moduli
action. Their presence hinges upon the details of the underlying
${\cal N}=(2,2)$  superconformal worldsheet theory.  These couplings
generalise known examples of the lifting of instanton
reparameterisation modulini through curvature couplings or
background fluxes.

Concerning this latter point, we tried to substantiate the well-motivated speculation \cite{Argurio:2007vq,Ibanez:2007rs,Aharony:2007pr}  that closed string background fluxes might also lift the universal $\ov \tau$ modes, restricting ourselves to the familiar framework of Type IIB orientifolds with supersymmetric three-form flux. In agreement with the results in particular of \cite{Bergshoeff:2005yp}, in the absence of gauge flux on the $E3$-instanton no such lifting occurs. We showed, building on the instanton action derived in \cite{Tripathy:2005hv}, that once worldvolume fluxes are turned, a lifting might be possible, but only in situations where the divisor wrapped by the instanton contains non-trivial two-cycles anti-invariant under the orientifold action. As it stands we have to leave it open whether this effect can actually be realised in explicit models and, if so, whether it enables the instanton to contribute to the superpotential. As one of the most imminent open questions it therefore remains to study a concrete global example in the spirit of the setup discussed in the last section. Also, it would be desirable to gain comparable understanding of the effects of Type IIA fluxes on the $E2$-instanton zero modes.

\vskip 1cm
 {\noindent  {\Large \bf Acknowledgements}}
 \vskip 0.5cm
\noindent We thank  N. Akerblom, V. Braun, K.S. Choi, R. Donagi, E.
Kiritsis, D. L\"ust, F. Marchesano, S. Moster, E. Plauschinn, M.
Schmidt-Sommerfeld, S. Stieberger, G. Shiu, D. Tong and A. Uranga
for interesting discussions as well as L. Dixon for helpful
correspondence. R.B. thanks the University of Pennsylvania for
hospitality. M.C., R.R. and T.W. are grateful to ASC and
Max-Planck-Institut f\"ur Physik, Munich, as well as to CERN for
hospitality. This research was supported in part by DOE grant
DOE-EY-76-02-3071 and the Fay R. and Eugene L. Langberg Endowed
Chair.

\newpage
\appendix

\section{Orientifold projection of instanton zero modes}
\label{orientifold}

In this appendix we describe explicitly the orientifold action $\Omega \ov \sigma$ on the zero modes of an $E2$-instanton wrapping the cycle $\Xi$. If $\Xi$ is not invariant under the orientifold action one includes, in the upstairs picture,  the orientifold image $E2'$ wrapping the image cycle $\Xi'$. The orientifold action identifies the $E2-E2$ modes with the $E2'-E2'$ modes and
$E2-E2'$-modes with $E2'-E2$-modes.
The $E2-E2'$ modes arising at invariant intersections on top of the orientifold plane are  symmetrised/anti-symmetrised as will be described momentarily. The same applies to the $E2-E2$ sector if the $E2$ wraps a cycle invariant under the orientifold action, $\Xi=\Xi'$.

The orientifold action on the bosonic and fermionic instanton zero modes in the invariant sector can be deduced from the action on spacetime-filling $D6$-branes wrapping the same internal cycle  $\Xi$ (and possibly its image)  as follows:
\begin{itemize}
\item[(1)]
The orientifold action on the internal oscillator part of the vertex operators agrees in the $D6$ and $E2$ case.
The only difference in the $E2$ case is that the external 4D space is orthogonal to the $E2$-brane and thus counts as transverse when applying the usual rules for representing $\Omega \ov \sigma$.
This entails the inclusion of an additional minus sign for bosonic excitations in the external 4D space and the inclusion of a factor  $e^{i \pi (s_0 + s_1)}$ for all fermionic zero modes. Here  $e^{i \pi (s_0 + s_1)}$ acts on the (anti-)chiral 4D spin fields $S^{\alpha}$ ($S^{\dot \alpha}$) as $ e^{i \pi (s_0 + s_1)} S^{\alpha} = -1$ ( $e^{i \pi (s_0 + s_1)} S^{\dot \alpha} = 1$).
\item[(2)]
Let $\gamma_{\Omega \ov \sigma,D6}$ denote the matrix representing the orientifold action on the CP factors of the $D6$-brane modes. Then the corresponding matrix for the $E2$-instanton $\gamma_{\Omega \ov \sigma,E2}$ enjoys
\bea
\label{gamma}
\gamma_{\Omega \ov \sigma,D6} = \pm \gamma_{\Omega \ov \sigma,D6}^T \Longleftrightarrow  \gamma_{\Omega \ov \sigma,E2} = \mp \gamma_{\Omega \ov \sigma,E2}^T.
\eea
The + and - cases for the projection relevant for D6-branes are referred to as orthogonal (SO) and symplectic (SP) projections, respectively, because for invariant D6-branes they yield gauge bosons in the adjoint of the respective gauge groups. In the latter case, invariant cycles have to be wrapped by an even number of D6-branes.

%The two cases correspond to putting the D6-branes on top of an $O^-$ and $O^+$-plane respectively. When we consider invariant intersections in D6-D6' sector, we have to take into account all O-planes present in the model to decide of which type the effective projection is. Our conventions will be such that for a single $O^+$-plane, also in the D6-D6' sector $\gamma_{\Omega \ov \sigma,D6} = -\gamma_{\Omega \ov \sigma,D6}^T$.

Finally, the relation (\ref{gamma}) follows via T-duality from the D9-D5 system analysed by Gimon-Polchinski \cite{Gimon:1996rq}.

\end{itemize}
It is straightforward to apply these rules to the zero modes for two different cases: (i)\, the universal zero modes for $\Pi_{\Xi} =\Pi_{\Xi'}$ and (ii) the modes in the $E2-E2'$ sector arising on top of the orientifold for $\Pi_{\Xi} \neq \Pi_{\Xi'}$.\\
In case the instanton wraps a cycle $\Pi_{\Xi} =\Pi_{\Xi'}$ the orientifold action on the universal zero modes $x^{\mu}$ and $\theta^{\alpha},\ov \tau^{\dot \alpha}$ leads to
\bea
&& \Omega_{x^\mu} =  \gamma_{E2} \Omega_{x^\mu}^T \gamma_{E2}^{-1}, \\
&& \Omega_{\theta^{\alpha}} =  \gamma_{E2} \Omega_{\theta^{\alpha}}^T \gamma_{E2}^{-1}, \quad\quad\quad \Omega_{\ov\tau^{\dot\alpha}} =  -\gamma_{E2} \Omega_{\ov\tau^{\dot\alpha}}^T \gamma_{E2}^{-1},
\eea
where for $x^{\mu}$ and $\theta^{\alpha}$ the minus sign due to the excitation gets cancelled by the minus sign due to rule (1).
Thus for a single instanton subject to the projection $\gamma_{E2} = \gamma_{E2}^T$, only  $x^{\mu}$ and $\theta^{\alpha}$ survive.\\
Modes in the $E2-E2'$-sector arising at intersection on top of the orientifold get (anti-)symmetrised as follows:

\noindent If for D6-branes wrapping the same cycle the invariant states get anti-symmetrised, then
\bea
&& \Omega_{m/ \ov m} =  \Omega_{m/ \ov m}^T , \quad\quad  \Omega_{\ov \mu} = \Omega_{\ov \mu}^T,\\
&& \Omega_{\mu} = - \Omega_{\mu}^T.
\eea
This results in the intersection numbers displayed in table \ref{antizero}.
In particular, for a single instanton, the
 zero mode $\mu^{\alpha}$ gets projected out and only $m, \ov m, \ov \mu^{\dot\alpha}$
 survive corresponding to case I in section \ref{sec_zero_U1}

If for D6-branes the invariant states are symmetrised, everything just changes sign.
 %with $\gamma_{E2} = - \gamma_{E2}^T$ only $\mu^{\alpha}$ survives, while if  $\gamma_{E2} = \gamma_{E2}^T$

\section{Details of the CFT computations}\label{closed_int}

In this appendix we demonstrate the computation of the amplitude
$\langle   m \, \ov \tau \,  \ov \mu\rangle$ as well as of some of the
couplings of the fermionic zero modes of the instanton to the closed
string background relevant for the F-term corrections investigated
in section \ref{sec_BW} and \ref{Kahler}. For simplicity we focus on
the case of an instanton wrapping a factorizable cycle of a toroidal
orbifold. More details of the CFT computation in this context can be found in \cite{Cvetic:2007ku}.
While  for other backgrounds the presence of the couplings in question has to be checked in concrete computations, all couplings which do not violate any of the general selection rules of the ${\cal N}=2$ SCFT on the worldsheet are generically present.

Let us start with the open string coupling $\langle   m \, \ov \tau
\,  \ov \mu \rangle$ used in eq. \ref{rec-actionb}. The relevant vertex operators take the
form\footnote{Here we assume the most symmetric configuration in which all
intersection angles $\theta^{i}_{E2E2'}>0$ and
$\sum^{3}_{i=1}\theta^{i}_{E2E2'}=2$.}
\begin{align}
V^{(-\frac{1}{2})}_{\bar{\tau}}(z)&=\Omega_{\bar{\tau}}\,\bar{\tau}^{\dot{\alpha}}\,
\,S_{\dot{\alpha}}(z)\,\prod^3_{i=1}\, e^{-i/2\,H_i(z)}
\,e^{-\varphi/2(z)}\nonumber  \\\label{ovtheta -1/2}
V^{(-1)}_{m}(z)&=\Omega_{m}\, m \, \prod^3_{i=1}\,
e^{i(1-\theta^i_{E2E2'})\,H_i(z)} \, e^{-\varphi(z)}\\ \nonumber
V^{(-\frac{1}{2})}_{\bar{\mu}}(z)&=\Omega_{\bar{\mu}}\,
\bar{\mu}^{\dot{\alpha}} \,S_{\dot{\alpha}}(z) \prod^3_{i=1}\,
e^{-i(\frac{1}{2}-\theta^i_{E2E2'})\,H_i(z)} \, e^{-\varphi(z)}.
\end{align}
Inserting them into $\langle   m \, \ov \tau \,  \ov \mu \rangle$
leads to
\begin{align*}
\langle   m \, \ov \tau \,  \ov \mu \rangle &= Tr(\Omega_m \,
\Omega_{\bar{\tau}} \, \Omega_{\bar{\mu}} )\,m \,
\bar{\tau}^{\dot{\alpha}}  \,
\bar{\mu}^{\dot{\beta}}\,<e^{-\varphi(z_1)}\, e^{\varphi/2(z_2)}\,
e^{-\varphi/2(z_3)}>\\& <S_{\dot{\alpha}}(z_1)\,
S_{\dot{\beta}}(z_3) > \, \prod^3_{i=1}
<e^{i(1-\theta^i_{E2E2'})\,H_i(z_1)}\,e^{-i/2\,H_i(z_2)} \,
e^{-i(\frac{1}{2}-\theta^i_{E2E2'})\,H_i(z)}>,
\end{align*}
which together with the supersymmetry condition
$\sum^{3}_{i=1}\theta^{i}_{E2E2'}=2$ results in the coupling \bea
m \,\bar{\tau}_{\dot{\alpha}} \,\bar{\mu}^{\dot{\alpha}}. \eea

Now we turn to interactions between fermionic zero modes of the
instanton and closed string background fields. We start with the
coupling between the reparametrization modulini $\bar{\chi}^{\dot \alpha}$
surviving the orientifold action and  the anti-chiral K\"ahler
modulini $\bar{t}^{\dot \alpha}$. Their vertex operator in type IIA takes the form
\begin{align} V^{(-\frac{1}{2})}_{\bar{\chi}^I}(z)
&=\Omega_{\bar{\chi}} \,\bar{\chi}^{I\,\dot{\alpha}}
\,\,e^{-\varphi/2(z)} \,
S_{\dot{\alpha}}(z) \, e^{-i/2 H_I(z)} \prod^3_{i\neq I} e^{i/2 H_i(z)},\\
 V^{(-\frac{1}{2},-1)}_{\bar{t}^{IJ}}(z)&= \bar{t}^{IJ\, \dot{\alpha}}
\,e^{-\varphi/2(z)}\,S_{\dot{\alpha}}(z)\,e^{-i/2H_I(z)}\,\prod^3_{i\neq
I}\, e^{i/2\,H_i(z)} \,e^{-\tilde{\varphi}(\bar{z})}
\,e^{-i\tilde{H}_J(\bar{z})}(\bar{z})\,
 e^{ikX(z,\bar{z})}.
 \end{align}
Note that on a factorizable torus $T^6 = T^2 \times T^2 \times T^2$
only the diagonal moduli $\bar{t}^{JJ}$ survive.

We see that the couplings $\langle \ov \chi \, \ov t\rangle$ respect
the total $U(1)$ worldsheet charge. However, only the amplitudes
$<\bar{\chi}^I\,\bar{t}^{JK}>$ for $I\neq J \neq K \neq I$ preserve
the internal $U(1)$-charge in each $T^2$ separately and lead to a
coupling \bea {\bar{\chi}^I}_{\dot{\alpha}}\,
\bar{t}^{JK\,\dot{\alpha}} \label{chi t}. \eea While on factorizable
tori $(T^2)^3$ and orbifolds thereof no such couplings exist, on
general Calabi-Yau threefolds there is no reason for them to vanish.

On the other hand, one
can convince oneself that the anti-chiral complex structure
modulini with vertex operators \bea
V^{(-\frac{1}{2},-1)}_{\bar{u}^{IJ}(z)} = \bar{u}^{IJ\,
\dot{\alpha}}
\,e^{-\varphi/2(z)}\,S_{\dot{\alpha}}(z)\,e^{-i/2H_I(z)}\,\prod^3_{i\neq
I}\, e^{i/2\,H_i(z)} \,e^{-\tilde{\varphi}(\bar{z})}
\,e^{i\tilde{H}_J(\bar{z})}(\bar{z})\,
 e^{ikX(z,\bar{z})}
 \label{U}
\eea
 do not couple to $\bar{\chi}$ due to non conversation of the total $U(1)$
world sheet charge. This is therefore a universal result.

 The corresponding bosonic superpartner terms to \eqref{chi t}
arise from amplitudes of the form \bea <\theta^{(+1/2)}
\bar{T}^{(-1,-1)} \bar{\chi}^{(-1/2)}
>\,\,,
\label{amplitude1}\eea where the superscripts denote ghost picture
of the respective vertex operator. Note that with the choice
displayed in \eqref{amplitude1} we ensure the total ghost charge
constraint. The vertex operator of the K\"ahler moduli takes the
form \begin{align*} V^{(-1,-1)}_{\bar{T}^{IJ}}= \bar{T}^{IJ}
e^{-\varphi(z)}\, e^{-iH_I(z)}\, e^{-\tilde{\varphi}(\bar{z})}\,
e^{-i\tilde{H}_{J}(\bar{z})}\,e^{ikX(z,\bar{z})}
\end{align*}
while the one for the $\theta$-mode in $(+\frac{1}{2})$-ghost
picture is given by
\begin{align}
V^{(+\frac{1}{2})}_{\theta^{\alpha}}(z)&=\Omega_{\theta}\,\theta^{\alpha}\,
\Big[\partial
X_{\mu}\,(\sigma^{\mu})^{\,\,\,\dot{\alpha}}_{\alpha\,}\,S_{\dot{\alpha}}(z)\,\prod^3_{i=1}\,
e^{i/2\,H_i(z)} \nonumber\\&\qquad \qquad +
\,\,\,\,\,\,\sum^3_{I=1}S_{\alpha}(z) \partial Z^{I}
e^{-i/2H_{I}(z)} \,\prod_{i \neq I} e^{i/2H_{i}(z)}\Big]
\,e^{\varphi/2(z)}. \label{theta 1/2}
\end{align}
By internal U(1) charge conservation only the first summand
contributes to the amplitude and in addition one has to require, as
for $<\bar{\chi}\, \bar{t}>$, that $I\neq J \neq K \neq I$. Then
\begin{align*}
<\theta\,\bar{T}^{JK} \, \bar{\chi}^{I}>&= Tr(\Omega_{\theta}\,
\Omega_{\bar{\chi}}) \,\,\theta^{\alpha}
(\sigma^{\mu})^{\,\,\,\dot{\alpha}}_{\alpha\,} \,
\bar{\chi}^{I\,\dot{\beta}} \, \bar{T}^{JK}\\ &<e^{\varphi/2(z_1)}\,
e^{-\varphi(z_2)}\, e^{-\tilde{\varphi}(\bar{z}_2)}\,
e^{-\varphi/2(z_3)}> <S_{\dot{\alpha}}(z_1)\, S_{\dot{\beta}}(z_3) >
\\& <e^{i/2H_I(z_1)}\, e^{-i/2H_I(z_3)}> <e^{i/2H_J(z_1)}\,e^{-iH_J(z_2)}
e^{i/2H_J(z_3)}>\\&<e^{i/2H_K(z_1)}\,e^{-i\tilde{H}_K(\bar{z}_2)}
e^{i/2H_K(z_3)}> <\partial X_{\mu}(z_1)  \, e^{ik_2X(z_2,\bar{z}_2)}
>.
\end{align*}
The correlators are easily evaluated and lead to couplings
proportional to
\begin{align*}
\theta \,\sigma^{\mu}\, \bar{\chi}^{I} \,\, \partial_{\mu}\,
\bar{T}^{JK}.
\end{align*}
As above, by non-conservation of $U(1)$-charge
there is no coupling to the bosonic  complex structure field
$U$. On the other hand the amplitude $<\theta u^{IJ}>$ is
non-vanishing. Here the vertex operator of $u^{IJ}$ is the complex
conjugated of \eqref{U}
\begin{align*}
V^{(-\frac{1}{2},-1)}_{u^{IJ}}(z) &= u^{IJ}_{ \alpha}
\,e^{-\varphi/2(z)}\,S^{\alpha}(z)\,e^{i/2H_I(z)}\,\prod^3_{i\neq
I}\, e^{-i/2\,H_i(z)} \,e^{-\tilde{\varphi}(\bar{z})}
\,e^{-i\tilde{H}_J(\bar{z})}(\bar{z})\,
 e^{ikX(z,\bar{z})}
\end{align*}
and the vertex operator for $\theta$ in $(-\frac{1}{2})$-ghost
picture takes the form
\begin{align}
V^{(-\frac{1}{2})}_{\theta}(z)&=\Omega_{\theta}\,\theta_{\alpha}\,
\,S^{\alpha}(z)\,\prod^3_{i=1}\, e^{i/2\,H_i(z)}
\,e^{-\varphi/2(z)}. \label{theta -1/2}
\end{align}
Now, one can easily check that U(1) world sheet charge is conserved
only in case $I=J$ and the resulting coupling takes the form \bea
\theta^{\alpha} \, u^{II}_{\alpha} \label{u1} \,\,.\eea

For the couplings relevant in section \ref{Kahler} we also need the
corresponding bosonic partner arise from amplitudes involving
$a^{IJ}$ and $\bar{\omega}^{IJ}$. For brevity we only display the
computation of the amplitude $<\theta^{(1/2)} \bar{\omega}^{(-1,-1)}
\ov\tau^{(-1/2)}>$, where the vertex operator for $\bar{\omega}$ is
\begin{align}
V^{(-1,-1)}_{\bar{\omega}^{IJ}}(z) = \bar{\omega}^{IJ}
e^{-\varphi(z)}\, e^{iH_I(z)}\, e^{-\tilde{\varphi}(\bar{z})}\,
e^{-i\tilde{H}_{J}(\bar{z})}\,e^{ikX(z,\bar{z})}\,\,,
\end{align}
while the vertex operator for the $\theta$ and $\bar{\tau}$ in the
respective ghost picture are given by \eqref{theta 1/2} and
\eqref{ovtheta -1/2}.
%\begin{align}
%V^{(-\frac{1}{2})}_{\bar{\tau}}(z)&=\Omega_{\bar{\tau}}\,\bar{\tau}^{\dot{\alpha}}\,
%\,S_{\dot{\alpha}}(z)\,\prod^3_{i=1}\, e^{-i/2\,H_i(z)}
%\,e^{-\varphi/2(z)}. \label{ovtheta -1/2}
%\end{align}
Again U(1) world sheet charge requires $I=J$ and a
computation analogous to the one leading to the amplitude $<\theta \, \bar{T} \,\bar{\chi}> $
gives the coupling \bea \theta \, \sigma^{\mu} \, \bar{\tau} \,\,
\partial_{\mu} \,\bar{\omega}^{II}. \eea
On the other hand due to the U(1) world-sheet charge there the
amplitudes $<\bar{\tau} \, \bar{t} >$ as well as $<\theta
\,\bar{T} \, \bar{\tau}>$ vanish.

\clearpage
\nocite{*}
\bibliography{rev}
\bibliographystyle{utphys}

\end{document}